\documentclass[fleqn, final,12pt]{elsarticle}
\usepackage[margin=1in]{geometry}

\usepackage{physics}
\usepackage{siunitx}
\usepackage{amsmath,amssymb}
\usepackage{anyfontsize}
\usepackage{enumitem}
\usepackage{cleveref}
\usepackage{booktabs,tabularx,makecell}
\usepackage{url}
\usepackage{threeparttable}

\def\tsc#1{\csdef{#1}{\textsc{\lowercase{#1}}\xspace}}
\tsc{WGM}
\tsc{QE}

\DeclareMathOperator*{\argmax}{arg\,max}
\DeclareMathOperator*{\argmin}{arg\,min}

\def\x{\mathbf{x}}

\journal{}

\begin{document}
\begin{frontmatter}

\title{A conservative micro-continuum-cellular automaton method for multispecies biofilm dynamics in complex flows}

\author[inst1]{Soyoung Kim}
\cortext[cor1]{Corresponding author}
\author[inst1]{Yinuo Noah Yao\corref{cor1}}

\affiliation[inst1]{organization={Zachry Department of Civil and Environmental Engineering},
            addressline={Texas A\&M University}, 
            city={College Station},
            postcode={77843}, 
            state={TX},
            country={USA}}

\begin{abstract}
We develop a conservative micro-continuum-cellular automaton method for simulating multispecies biofilm dynamics in complex flows. The proposed method couples the Darcy-Brinkman-Stokes equations, reactive transport, suspended bacteria, and biofilm dynamics with a two-stage cellular automaton algorithm for biofilm redistribution and interface evolution. While treating biofilms as evolving porous media, we ensure conservative redistribution of multispecies biomass across partially occupied cells. Donor and recipient cell volumes are explicitly accounted for to conserve biomass and preserve species composition on non-uniform meshes. The proposed method is assessed against diffusion-dominated benchmark cases, including single-species fingering and multispecies stratification, and is further evaluated through a mesh-convergence study for flow and growth over a rectangular bump. The framework is then applied to counter-diffusional biofilms in a membrane-aerated biofilm reactor as a canonical example. The results demonstrate that the framework captures the expected biofilm morphology and stratification in systems involving coupled flow, substrate transport, and biofilm dynamics. The proposed method provides a flexible computational approach for simulating multispecies biofilm dynamics in complex flows and geometries.
\end{abstract}

\end{frontmatter}

\section{Introduction}
\label{intro}

Biofilm dynamics due to the coupling between flow, substrate transport, and microbial activity are important in a wide range of natural systems  (e.g., sediment transport, natural wetlands, and hyporheic exchange) \cite{Xian2022,Xian2019-lo, Jasper2014-oo,Fang2017-yr} and engineered systems (e.g., water and wastewater treatment, and marine structures) \cite{Feng2013, Yuan2020, Lindholdt2015}. Biofilms typically exist as multispecies systems, in which multiple microbial species compete for numerous substrates \cite{Ghoul2016}. Biofilm dynamics are also controlled by shear-induced detachment, which continuously reshape biofilms and regulate their distribution under flow conditions \cite{Baveye1998, Horn2003}. Through growth and accumulation, biofilms can significantly alter system behavior and performance by modifying local substrate transport, flow pathways, and hydraulic resistance \cite{Kurz2022, Kurz2023, Yan2017, Deng2013, Vandevivere1992}. For example, in hyporheic exchange between surface water and groundwater, biofilm accumulation in the soil can lead to pore clogging, thereby reducing permeability and limiting water exchange \cite{Xian2019-lo}. Similarly, in biofilters used for water treatment, the balance between biofilm growth and hydraulic conductivity is critical to system performance and operational stability \cite{Le-Bihan2000-ze,Jeong2017-br}.

However, given the challenges of predicting biofilms in various flow environments and experimentally quantifying biofilm distributions \cite{Mattei2017}, multiple approaches have been developed to model \cite{DAcunto2019, Eberl2001, Picioreanu1998a, Wanner1986} or numerically simulate biofilm dynamics \cite{Li2025,Benioug2017,Tang2013-dd, Jung2021} in complex domains. In contrast to typical dissolution/precipitation problems, where the fluid-solid interface evolves through interfacial reactions \cite{Yousefzadeh2023-rg, Li2010-ro, Li2008-bs}, biofilms evolve through both interfacial movement and internal growth/decay dynamics: substrates are transported into permeable biofilms, leading to the growth of biomass within the porous matrix. As a result, the accumulated biomass displaces the fluid-biofilm interface. Based on the representation of biofilm movement, existing methods can be broadly classified into continuum \cite{Wanner1986, Alpkvista2007} and discrete approaches \cite{Picioreanu1998b, Tang2013}.

Most continuum approaches focus on low-Reynolds-number flow and represent biofilm dynamics using modified diffusion or advection equations. In these models, both effective velocity and/or diffusivity of biofilms require closure relations, such as relating them to a biofilm pressure field or assuming that biofilm movement follows potential flow \cite{Wanner1986, Rahman2015, Wood2002, Golfier2009, Klapper2002, Alpkvista2007}. Interactions between different microbial species within the biofilm can be modeled using cross-diffusion formulations~\cite{Rahman2015}. However, the validity of these closure relations, their extension to systems with multiple interacting species, and their applicability to moderate-Reynolds-number flows remain uncertain.

In contrast, discrete models such as cellular automaton (CA), which have been applied to single- and multispecies biofilms, simulate biofilm movement by redistributing excess local biomass to neighboring cells \cite{Tang2013, Rahman2015, Cockx2024, Li2025, Pizarro2001, Noguera2004}. Early CA models typically used fully local stochastic neighbor-to-neighbor displacement rules, which preserve locality but can result in spatial discontinuities in multispecies biofilms \cite{Picioreanu1998a, Picioreanu1998b, Kreft2001}. To reduce stochastic variability, later approaches modified the redistribution rules by allowing excess biomass to move along the shortest path to the biofilm-fluid interface through non-adjacent cells \cite{Noguera1999, Knutson2005, Laspidou2010}. Although these modifications improve spatial continuity, they can introduce artificial mixing during multispecies redistribution. Tang and Valocchi \cite{Tang2013} reduced this problem by modifying the redistribution procedure to better preserve local species structure.

In addition to biofilm movement, another major challenge in simulating biofilm dynamics is the coupling between fluid flow and biofilms. Most CA approaches assign each computational cell a binary state, either fluid or biofilm, once the maximum biomass threshold is reached \cite{Tang2013}. This leads to three limitations: (1) velocity fields are inaccurately predicted in partially filled cells, which are treated as fluid, (2) no-slip boundary conditions imposed at the biofilm-fluid interface result in diffusion-dominated transport within biofilms, and (3) partial voids within biofilms are disconnected from the main free flow field. However, numerous studies have shown that biofilms are heterogeneous and consist of loosely structured microbial aggregates containing patchy voids and channels~\cite{Thullner2008, Deng2013, Xian2022}. Flow pathways within biofilms have also been observed experimentally~\cite{Stoodley1994}, indicating that biofilms are permeable structures with permeabilities ranging from approximately $10^{-15}$ [\si{\meter\squared}] to $10^{-7}$ [\si{\meter\squared}] \cite{Deng2013, Dreszer2013, Karimifard2021, Kurz2023}. These approximations can also alter local substrate transport by changing the relative contributions of advection and diffusion within and around the biofilm.

Recently, micro-continuum approaches have been widely adopted to directly couple fluid flow in free-flow regions and microporous regions~\cite{Golfier2015, Soulaine2018, Guo2018, Carrillo2022} in reactive transport. This approach is formulated based on the Darcy-Brinkman-Stokes (DBS) equations, which combine free-flow regimes (Navier-Stokes) and porous regimes (Darcy) by incorporating viscous effects through the Brinkman term~\cite{Soulaine2016}. To date, most DBS-based approaches~\cite{Guibert2025, Li2025} for predicting biofilm dynamics have been coupled with continuum models of biofilm movement based on the effective-diffusivity formulation of \cite{Eberl2001}. In parallel, Brinkman-type flow has previously been coupled with CA-based biofilm growth \cite{Kapellos2007-qs}. Kapellos et al.~\cite{Kapellos2007-qs} combined Brinkman-type flow, reactive transport, and discrete biofilm spreading in porous media. However, conservative redistribution of multispecies biomass between partially occupied computational cells on non-uniform meshes remains less developed.

In this work, we develop a conservative micro-continuum-CA method to simulate multispecies biofilm dynamics in complex flows and geometries. We extend the CA approach to non-uniform meshes by accounting for differences in cell volume during biomass redistribution while preserving species composition. We further refine the spreading rule to combine stochastic local expansion with directional interface advancement while maintaining consistent coupling with the DBS framework. The paper is organized as follows. In \Cref{methods}, we present the governing equations and the algorithm underlying the improved CA approach. In \Cref{verification}, we consider three verification cases to assess the accuracy of the proposed framework. In \Cref{sec:counter-biofilms}, we apply the framework to counter-diffusional multispecies biofilms, which require strong coupling between fluid flow, substrate transport, and biofilm dynamics.

\section{Numerical methodology}
\label{methods}

We develop a general computational framework to simulate the fine-scale coupling between hydrodynamics and bacterial processes. In this framework, bacteria are categorized into two types: (1) \textit{suspended bacteria}, which move freely with the fluid, and (2) \textit{attached bacteria} or ``biofilms,'' which represent clusters of immobilized bacterial cells. Both suspended and attached bacteria undergo growth and decay, with the coupling between the two phases by attachment and detachment processes (Figure~\ref{fig:biofilm_development}). The fluid flow in both the free-flow region and the biofilms is described by the modified DBS equation. Suspended bacteria are governed using modified scalar transport equations, whereas biofilms are described by coupling reactive equations with a CA algorithm to capture growth and morphological movement.

\subsection{Fluid flow in free-flow and biofilm regions}

\begin{figure}
  \centering
    \centerline{
        \includegraphics[scale=0.5]{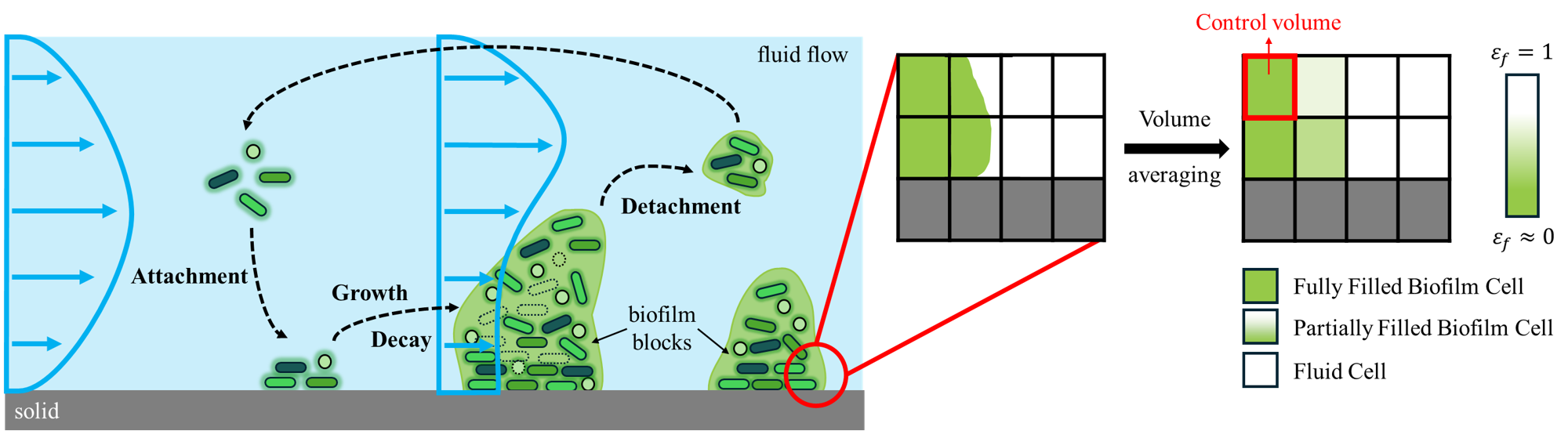}} 
    \caption{Schematic of a representative biofilm simulation. Fluid flow is shown by blue arrows, and the biofilm is represented as an evolving porous medium (green). Biofilm dynamics include attachment, detachment, growth, and decay, while suspended biomass is transported by advection and diffusion. The enlarged panels illustrate the volume-averaging procedure used to classify computational cells by the local porosity $\varepsilon_\text{f}$.}
    \label{fig:biofilm_development}
\end{figure}

To capture the interactions between biofilms and fluid flow, we treat biofilms as porous media that evolve over time. In both the free-flow and biofilm regions, the fluid flow is governed by modified DBS equations, which are widely used in subsurface reactive transport \cite{Soulaine2016,Vafai1981,Hsu1990, Bousquet2002, Goyeau2003}. In this approach (\Cref{fig:biofilm_development}), we consider each cell in the computational domain containing both pore fluid and biofilms (solid matrix). As such, porosity $\varepsilon_{\text{f}} \in [0, 1)$ is given by

\begin{align}
\label{eq:vf}
   \varepsilon_{\text{f}} = \frac{V_{\text{f}}}{V} = 1 - \sum_{i=1}^{N_{\text{b}}} \varepsilon_{\text{b},i},
\end{align}

\noindent where $V_{\text{f}}$ and $V$ are the fluid and total volumes of each computational cell, respectively, $N_{\text{b}}$ is the number of species considered in the biofilms, and $\varepsilon_{\text{b},i}$ is the volume fraction of species $i$. Here, we define the volume-average concentration of the suspended bacteria species $i$ ($\overline{X}_{i}^{\text{sb}}$ [\si{\kilogram\per\cubic\meter}]) and biofilm species $i$ ($\overline{X}_{i}^{\text{b}}$ [\si{\kilogram\per\cubic\meter}]) in the pore space as 

\begin{align}
    \overline{X}_{i}^{\text{sb}} = \frac{1}{V_{\text{f}}} \int_{V_{\text{f}}} X_{i}^{\text{sb}} \,\dd V , 
    \quad \overline{X}_{i}^{\text{b}} &= \frac{1}{V_{\text{f}}} \int_{V_{\text{f}}} X_{i}^{\text{b}}\,\dd V,
\end{align}

\noindent respectively. The volume fraction of biofilm species $i$ is then calculated as 

\begin{gather}
    \varepsilon_{\text{b}, i} = \frac{\varepsilon_{\text{f}}\overline{X}_{i}^{\text{b}}}{\rho_{\text{b}}^0},
\end{gather}

\noindent where $\rho_{\text{b}}^0$ [\si{\kilogram\per\cubic\meter}] is the dry density of the biofilm. We emphasize that $\rho_{\text{b}}^0$ is the theoretical combined density of cells and extracellular polymeric substances (EPS) that fully occupy a given volume. Since biofilms contain internal pores, the maximum biofilm density is typically lower than the dry density. Previous studies have reported volume fraction of biofilms ranging from 0.07 to 0.80~\cite{Kurz2023, Jung2021, Lewandowski1999, Deng2013, Karimifard2021, Zhang1994}, and a typical biofilm density varying between 10 and 100 [\si{\kilogram\per\cubic\meter}]~\cite{Benioug2017, Alpkvista2007, Zang2025, Zhang1994}. We therefore define the maximum biofilm density as $M_{\text{max}} = (1 - \varepsilon_{\mathrm{f},\min}) \rho_{\mathrm{b}}^{0}$, where $\varepsilon_{\mathrm{f},\min}$ is the minimum  porosity.

Following the definition of volume-average concentrations of suspended bacteria and biofilms, we define the average pressure $\overline{p}_{\text{f}}$ [\si{\pascal}] in the pore space as 

\begin{gather}
    \overline{p}_{\text{f}} = \frac{1}{V_{\text{f}}} \int_{V_{\text{f}}} p_{\text{f}} \,\dd V, 
\end{gather} 

\noindent whereas the average fluid velocity $\mathbf{\overline{u}_{\mathrm{f}}}~[\si{\meter\per\second}]$ is defined analogously to the velocity in Darcy's flow such that 

\begin{gather}
    \mathbf{\overline{u}_{\text{f}}} =\frac{1}{V} \int_{V_{\text{f}}} \mathbf{u_{\text{f}}} \, \dd V.
\end{gather}

\noindent The continuity and momentum equations using the DBS approach are then given by 

\begin{subequations}
    \label{eq:dbs_eqs}
    \begin{gather}
        \div{\mathbf{\overline{u}_{\text{f}}}} = 0  \label{eq:dbs_eq1} \\
        \frac{\rho_{\text{f}}}{\varepsilon_{\text{f}}}
        \left(\pdv{\mathbf{\overline{u}_{\text{f}}}}{t} + \div{\left(\frac{1}{{\varepsilon_{\text{f}}}}\mathbf{\overline{u}_{\text{f}}}\mathbf{\overline{u}_{\text{f}}}\right)}\right)
        = -\grad{\overline{p}_{\text{f}}} + \frac{\mu_\text{f}}{\varepsilon_{\text{f}}}{\grad}^2{\mathbf{\overline{u}_{\text{f}}}}-\mu_\text{f} k^{-1}\mathbf{\overline{u}_{\text{f}}} \label{eq:dbs_eq2}
    \end{gather}    
\end{subequations}

\noindent where $\rho_{\text{f}}$ [\si{\kilogram\per\cubic\meter}] is the fluid density, $\mu_\text{f}$~[\si{\kilogram\per\meter\per\second}] is the fluid dynamic viscosity, and $k$~[\si{\meter\squared}] is the permeability of the porous regions (biofilms in this study). The last term on the right side, $\mu_\text{f} k^{-1}\mathbf{\bar{u}_{\text{f}}}$, represents the momentum exchange term between the fluid and the biofilm phase (i.e. the Darcy resistance term). In typical porous media, $k$ is modeled as a function of $\varepsilon_{\text{f}}$ using relationships such as power law \cite{bear2013dynamics,Zhu1995,Hommel2018} and Kozeny-Carman \cite{kozeny1927, carman1937, Minto2019, Soulaine2018}. In this framework, we use the widely adopted Kozeny-Carman relationship with a structural parameter $k_0$, which in this case represents the intrinsic permeability of the biofilm. The equation is given by

\begin{align}
\label{eq:KC}
    k^{-1} = k_0^{-1} \frac{(1-\varepsilon_{\text{f}})^2}{\varepsilon_{\text{f}}^3}.
\end{align}

\noindent One advantage of Equations \eqref{eq:dbs_eqs} is that the momentum equation (\Cref{eq:dbs_eq2}) reduces to the Navier-Stokes equations with $\varepsilon_{\text{f}} = 1$, and approaches to Darcy's flow as $\varepsilon_{\text{f}} \rightarrow 0$. Equations \eqref{eq:dbs_eqs} are solved using the Pressure-Implicit with Splitting of Operators (PISO) \cite{ISSA1986} in OpenFOAM.

\subsection{Substrate transport and reaction}

In a typical biofilm-flow system, chemical substrates are transported through advection and diffusion in the free-flow region and within the biofilm, and are simultaneously consumed or produced by suspended bacteria and biofilms. Here, we consider the transport and reaction of $N_c$ chemical substrates that are transported from the free-flow region to the biofilm, where they are utilized by both suspended bacteria and biofilms. Transport is modeled using advection-diffusion equations, while reactions are described using widely adopted multiplicative Monod kinetics. In this study, we make two primary assumptions: (1) the utilization of substrate $j$ depends on its own concentration, the concentrations of other relevant substrates, and the concentration of suspended bacteria and biofilms, and (2) no inhibition occurs between different substrates. Additional microbiological processes can be easily incorporated by modifying the reaction term accordingly (e.g.~\cite{Lardon2011, Mehrani2021, Wang2022, Wang2020}). Similarly to the previous section, the average concentration of substrate $j$, $\overline{C}_{j}$~[\si{\kilogram\per\cubic\meter}], is defined as 

\begin{gather}
    \overline{C}_{j} =  \frac{1}{V_{\text{f}}} \int_{V_{\text{f}}} C_{j} \,\dd V.
\end{gather}

\noindent The governing equations of substrate $j$ are then  defined as

\begin{subequations}
\label{eq:substrates}
\begin{align}
\label{eq:adv_dif1}
\frac{\partial (\varepsilon_{\text{f}} \overline{C}_{j})}{\partial t}
+ \nabla \cdot (\overline{\text{u}}_{\text{f}}\, \overline{C}_{j})
- \nabla \cdot (\varepsilon_{\text{f}} D^*_{j} \nabla \overline{C}_{j}) = - \sum_{i=1}^{N_{\text{b}}} \alpha_{i,j} \varepsilon_{\text{f}}\left( \overline{X}_{i}^{\text{sb}} + \overline{X}_{i}^{\text{b}}\right)r^{\text{g}}_{i}, 
\end{align} 
\begin{align}
\label{eq:monod_1}
r^{\text{g}}_{i}=   \hat{q}^{\max}_{i} \prod_{k=1}^{N_{c}} \left(\frac{\overline{C}_k}{K_{i,k} + \overline{C}_k}
    \right)^{\beta_{i,k}}, 
\end{align}
\end{subequations}
\noindent

\noindent where $\overline{C}_{j}$ and $\overline{C}_{k}$ are the average concentrations of substrates $j$ and $k$, respectively, $D^*_{j}$~[\si{\meter\squared\per\second}] is the effective diffusion coefficient, $r^{\text{g}}_{i}$~[\si{\kilogram\per\cubic\meter\per\second}] is the substrate utilization rate by suspended bacterial or biofilm species $i$, $\hat{q}^{\max}_{i}$~[\si{\per\second}] is the maximum substrate utilization rate by species $i$, $K_{i,k}$~[\si{\kilogram\per\cubic\meter}] is the half-saturation coefficient of substrate $k$ by species $i$, $\beta_{i,k} = [0,1]$ is a constant indicating whether substrate $k$ is utilized by species $i$, and $\alpha_{i,j}$ is the stoichiometric coefficient of substrate $j$ in species $i$, with its sign indicating consumption ($\alpha_{i,j}<0$) or production ($\alpha_{i,j}>0$).

\subsection{Suspended bacteria and biofilm dynamics}

In this section, we focus on describing the governing equations of suspended bacteria and biofilms. The main difference between suspended bacteria and biofilms is their transport. We model suspended bacteria analogously to passive scalars, while adding source terms for growth, decay, attachment, and detachment. In contrast, processes such as reaction and detachment within the biofilm are treated in continuous form, whereas biofilm movement is governed by the discrete CA approach described in \Cref{sec:spreading-CA}. The governing equations for species $i$ of suspended bacteria and biofilms are given by

\begin{subequations}
\label{eq:bacteria}
\begin{align}\label{eq:suspended}
    \frac{\partial \left(\varepsilon_{\text{f}}\overline{X}_{i}^{\text{sb}}\right)}{\partial t} + \nabla \mathbf{\cdot} (\overline{\mathbf{u}}_{\text{f}}\overline{X}_{i}^{\text{sb}}) - \nabla \mathbf{\cdot} (\varepsilon_{\text{f}}D^*_{i} \nabla \overline{X}_{i}^{\text{sb}}) = \mu^{\text{sb,g}}_{i} + \mu^{\text{sb,d}}_{i} - \mu^{\text{att}}_{i} + \mu^{\text{det}}_{i},
\end{align}
\begin{align}\label{eq:biofilms}
    \dv{\left(\varepsilon_{\text{f}}\overline{X}_{i}^{\text{b}}\right)}{t} 
    = \mu^{\text{b,g}}_{i} + \mu^{\text{b,d}}_{i} + \mu^{\text{att}}_{i} - \mu^{\text{det}}_{i},
\end{align}
\end{subequations}

\noindent where $D^*_{i}$~[\si{\meter\squared\per\second}] is the effective diffusion coefficient of the species $i$, $m = \{ \text{sb}, \text{b}\}$ refers to suspended bacteria and biofilms, respectively, $\mu^{\text{m,g}}_{i}$ [\si{\kilogram\per\cubic\meter\per\second}] is the growth rate of species $i$, $\mu^{\text{m,d}}_{i}$ [\si{\kilogram\per\cubic\meter\per\second}] is the decay rate of species $i$, and $\mu^{\text{att}}_{i}$ [\si{\kilogram\per\cubic\meter\per\second}] and $\mu^{\text{det}}_{i}$ [\si{\kilogram\per\cubic\meter\per\second}] are the respective attachment and detachment rates of species $i$. The terms on the right-hand sides of \Cref{eq:suspended,eq:biofilms} represent the biological processes governing the dynamics of suspended bacteria and biofilms, and are summarized as follows:

\begin{description}
\item(1) \textbf{Growth and decay}: When suspended bacteria or biofilms consume substrates, part of the released energy is used for maintenance and the remainder is for biomass synthesis. Growth is therefore taken to be proportional to substrate consumption through the yield coefficient $Y_i$, defined as the mass of species $i$ produced per unit mass of substrate consumed. The growth rates of suspended bacteria and biofilms are then computed as

\begin{subequations}
\label{eq:growth}
\begin{align}
\label{eq:g1}
    \mu^{\text{sb},g}_{i} &= \varepsilon_{\text{f}}Y_{i}  r^{\text{g}}_{i} \overline{X}^{\text{sb}}_{i}, \\
    \mu^{\text{b},g}_{i} &= \varepsilon_{\text{f}} Y_{i}  r^{\text{g}}_{i} \overline{X}^{\text{b}}_{i}.
\end{align}
\end{subequations}

Regarding the decay rates, we model them as first-order reactions with a decay constant $b^{\text{d}}_{i}$ such that

\begin{subequations}
\begin{align}
\label{eq:decay}
    \mu^{\text{sb, d}}_{i} &= \varepsilon_{\text{f}}b^{\text{d}}_{i} \overline{X}^{\text{sb}}_{i} \\  
    \mu^{\text{b, d}}_{i} &= \varepsilon_{\text{f}}b^{\text{d}}_{i} \overline{X}^{\text{b}}_{i}.
\end{align}
\end{subequations}

\item(2) \textbf{Attachment and detachment}: The coupling between suspended bacteria and biofilms is through the process of attachment and detachment. In a typical system, suspended bacteria attach to existing biofilms and other surfaces to form new biofilms, whereas biofilms detach and become suspended bacteria. Here, we model these two processes as a first-order reversible reaction such that 

\begin{align}
    \overline{X}^{\text{sb}}_{i} \;\overset{\text{attach}}{\underset{\text{detach}}{\rightleftharpoons}}\; \overline{X}^{\text{b}}_{i}. 
\end{align}

For attachment, the rate of attachment for species $i$ is defined as

\begin{align}
\label{eq:attach}
    \mu^{\text{att}}_{i} &= \varepsilon_{\text{f}}k^{\text{att}}U^{\text{att}}\overline{X}^{\text{sb}}_{i}, 
\end{align}

\noindent where $k^{\text{att}}$ is the maximum rate of attachment and  $U^{\text{att}}$ refers to the success rate of attachment or the probability of successful attachment. Ref \cite{Guanju2023} has shown that attachment is highly dependent on the magnitude of the local velocity. For example, \emph{Pseudomonas putida} can only develop biofilms when the local velocity is less than $\num{5e-5}$ \unit{\metre\per\second}. Following \cite{Minto2019}, we model the probability of successful attachment using a stepwise function based on local velocity. No attachment would occur when the magnitude of the local velocity ($U_{\text{int}}$) is above the upper critical velocity ($U_{U}$), while attachment occurs at the maximum rate below the lower critical velocity ($U_{L}$). For $U_L \le U_{\text{int}} \le U_U$, the attachment rate varies linearly with $U_{\text{int}}$. The equation is given by 

\begin{align}
    U_{\text{att}} =
    \begin{cases}
    1, & \text{if } U_{\text{int}} < U_{L},\\[4pt]
    1 - \dfrac{\lvert U_{\text{int}}\rvert - U_{L}}{\,U_{U} - U_{L}\,}, & \text{if } U_{L} \le U_{\text{int}} \le U_{U},\\[8pt]
    0, & \text{if } U_{\text{int}} > U_{U}.
\end{cases}
\end{align}

Regarding biofilm detachment, we assume that the only mechanism of detachment is due to shear stresses acting on the biofilm-fluid interface. Following similar expression as attachment, the rate of biofilm detachment is given by

\begin{align}
\label{eq:detach}
    \mu^{\text{det}}_{i} &= \varepsilon_{\text{f}} k^{\text{det}}\tau^{\text{det}} \overline{X}^{\text{b}}_{i},
\end{align}    

\noindent where $k^{\text{det}}$ is the maximum rate of detachment and $\tau^{\text{det}}$ refers to the probability of detachment. Past studies have shown that detachment due to shear stress occurs only when the local shear stress is above a threshold value \cite{Choi2003, Stoodley2001, Sudarsan2005}. Here, we model the probability of detachment as a function of local shear stress $\tau$ and threshold shear stress $\tau_{\text{max}}$ \cite{Benioug2017} such that

\begin{align}
    \tau_{\text{det}} =
    \begin{cases}
    0, & \text{if } \tau < \tau_{\max},\\[4pt]
    \exp\!\left(\dfrac{\tau - 2\,\tau_{\max}}{\tau_{\max}} \right),
    & \text{if } \tau_{\max} \le \tau \le 2\,\tau_{\max},\\[8pt]
    1, & \text{if } \tau \ge 2\,\tau_{\max}.
\end{cases}
\end{align}

\noindent The present formulation uses hydrodynamic closures for attachment and detachment and does not explicitly resolve species-dependent adhesion or EPS-mediated cohesion.
\end{description}

\subsection{Cellular automaton algorithm for biomass redistribution}
\label{sec:spreading-CA}

\begin{figure}
  \centering
    \centerline{
        \includegraphics[scale=0.9]{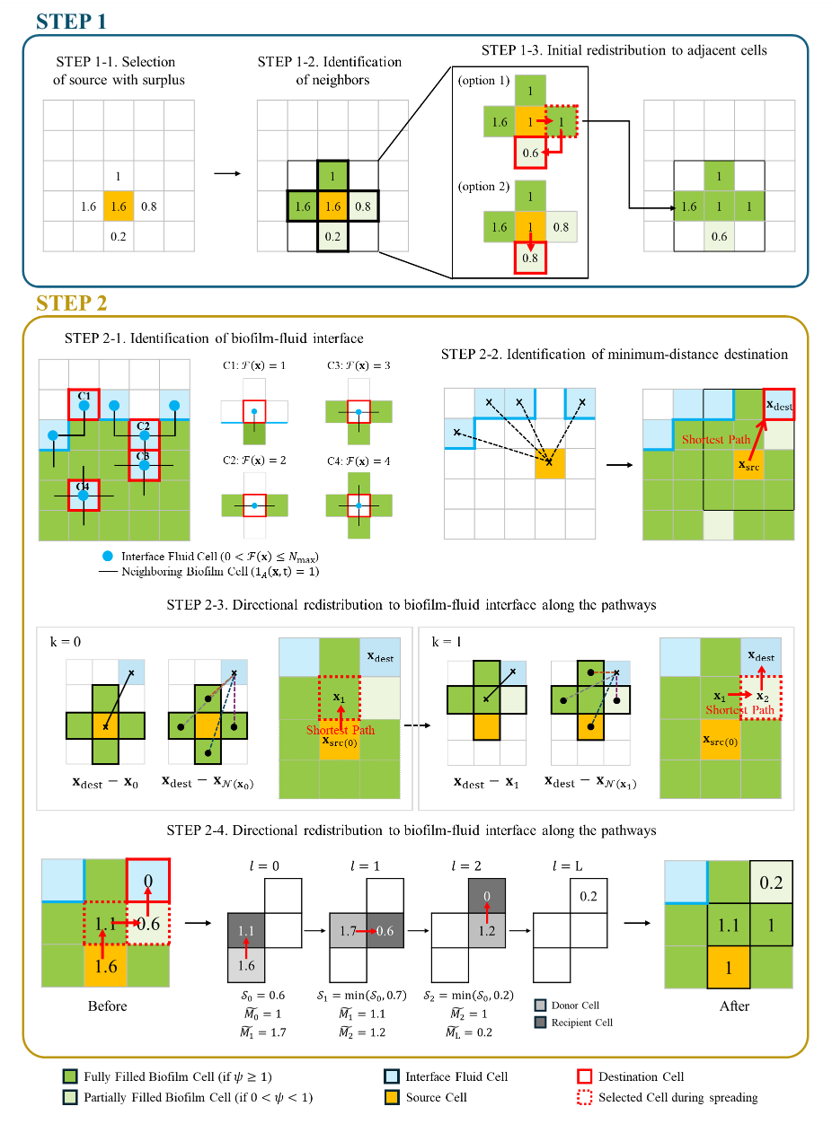}} 
    \caption{Two-stage biofilm spreading algorithm. Cell values are the cell state $\psi$. Stage 1 redistributes excess biomass to neighboring cells, and Stage 2 directs the remaining excess toward biofilm-fluid interface cells along minimum-resistance pathways.}
    \label{fig:CA_app1}
\end{figure}

In the previous section, we described the local biofilm dynamics (e.g., growth, attachment, and detachment), whereas in this section, we focus on the movement of the biofilm-fluid interface. Given the stochastic nature of interfacial expansion, we utilize discrete CA approaches to advance the biofilm-fluid interface. In this approach, when the local biofilm density exceeds the prescribed threshold ($M_{\max}$), the excess biofilm is redistributed in a single time step through a two-stage process (Figure~\ref{fig:CA_app1}): (1) initial stochastic redistribution to neighboring cells (STEP 1), followed by (2) directional spread from the source to the biofilm-fluid interface with the least resistance (STEP 2). The CA redistribution represents growth-induced biomass displacement rather than mechanical deformation of the biofilm. Biofilm compression, elasticity, and viscoelastic deformation are not explicitly resolved in the present formulation.

Redistribution is initiated by comparing the local biofilm density with the maximum biomass density, $M_{\max}$. The local biofilm density in a cell ($\overline{M}$) is computed as  

\begin{gather}
    \overline{M}=\varepsilon_{\text{f}} \sum_{i=1}^{N_{\text{b}}}  \overline{X}^{\text{b}}_{i}.
\end{gather}

\noindent When $\overline{M} \ge M_{\max}$, the excess biofilm will be redistributed following a defined pathway stochastically. This algorithm consists of three main components: (1) determining the state of each cell by tracking its current biofilm density and identifying whether excess biofilm exists, (2) identifying the spreading path of the excess biofilm, and (3) updating the biofilm density and species fractions along the spreading path until the excess is zero.\\

\noindent \textbf{Determining the cell states}: We compute the current state of each cell $\psi$ as

\begin{gather}
\label{eq:cell_state}
    \psi (\x, t) = \frac{\overline{M} (\x, t)}{M_{\max}}. 
\end{gather}

\noindent For $\psi = 0$, the cell is treated as an empty (fluid) cell. For $0 < \psi < 1$, the cell is considered partially filled with biofilms. For $\psi = 1$, the cell is fully occupied with biofilm. If $\psi > 1$, the cell contains excess biofilm and spreading is required. In this case, the excess in the cell ($\mathcal{S}(\mathbf{x}, t)$) is computed as

\begin{align}
    \mathcal{S}(\mathbf{x}, t) = \overline{M}(\mathbf{x}, t) - M_{\max}.
\end{align}


\noindent \textbf{Redistributing excess biofilm stochastically (STEP 1)}: For the initial redistribution to neighboring cells (four neighbors in two dimensions and six in three dimensions, first spreading, STEP 1 in \Cref{fig:CA_app1}), we first classify each neighbor as fluid ($\psi = 0$), partially filled ($0 < \psi < 1$), or fully filled ($\psi = 1$). If there is at least one fluid neighbor, the excess biofilm is transferred to a randomly selected fluid cell, which then becomes partially filled. If there are no fluid cells present, the excess is redistributed to a randomly selected partially-filled cell instead (STEP 1-3 in \Cref{fig:CA_app1}). Any remaining excess is then iteratively redistributed to the next eligible neighbors until no excess remains. After each redistribution step, the states of the source and recipient cells are updated according to \Cref{eq:cell_state}. \\

\noindent \textbf{Identifying interfacial cells (STEP 2-1)}: Once the initial redistribution is complete, if $\mathcal{S} > 0$, we will perform directional spreading (second spreading, STEP 2 in \Cref{fig:CA_app1}). Before spreading, we first identify the biofilm-fluid interface (\Cref{fig:CA_app1} STEP 2-1) by examining each fluid cell ($\psi = 0$) and its neighboring cells such that

\begin{subequations}
\begin{gather}
    \mathcal{F} (\x, t) = \sum_{\mathbf{y} \in \mathcal{N}(\x)} \mathbf{1}_A (\mathbf{y},t)
\end{gather}

\noindent where $\mathcal{N}(\x)$ refers to the set of neighboring cells of the cell $\x$, and the indicator function $\mathbf{1}_A (\x,t)$ is given by

\begin{align}
    \mathbf{1}_A (\x,t) &= 
    \begin{cases}
        0, \quad \text{if } \psi = 0, \\
        1, \quad \text{if } \psi > 0.
    \end{cases}
\end{align}
\end{subequations}

\noindent In a Cartesian mesh, $\mathcal{F}(\mathbf{x}, t)$ ranges from 0 to $N_{\max}$, where $N_{\max}=4$ and 6 for two- and three-dimensional simulations, respectively. A cell is identified as a biofilm–fluid interfacial cell when $0 < \mathcal{F}(\mathbf{x}, t) \leq N_{\max}$. \\


\noindent \textbf{Determining the destination cell (STEP 2-2)}: Upon identification of biofilm-fluid interfacial cells, we determine the reference interfacial cell $\mathbf{x}_{\text{dest}}$ that minimizes the distance to the biofilm cell source as

\begin{gather}
    \mathbf{x}_{\text{dest}} = \argmin_{\mathbf{x}_i \in \Gamma_{\text{int}}} \|\mathbf{x}_i - \mathbf{x}_{\text{src}}\|_2 
\end{gather}

\noindent where $\Gamma_{\text{int}}$ is the set of all interfacial cells and $\mathbf{x}_{\text{src}}$ refers to the position of the source cell. If multiple interfacial cells have the same distance, we select one at random. \\

\noindent \textbf{Constructing the spreading path (STEP 2-3)}: Once the destination interfacial cell $\mathbf{x}_{\mathrm{dest}}$ is identified, the spreading path is constructed sequentially from the source cell $\mathbf{x}_{\mathrm{src}}$. Here, $\mathbf{x}_k$ is defined as the current cell along the path, with $\mathbf{x}_0=\mathbf{x}_{\mathrm{src}}$ at $k=0$. The next cell $\mathbf{x}_{k+1}$ is then selected from the set of neighboring cells $\mathcal{N}(\mathbf{x}_k)$ such that the distance from the neighbor to $\mathbf{x}{\mathrm{dest}}$ is minimized relative to the current distance between $\mathbf{x}_k$ and $\mathbf{x}_{\mathrm{dest}}$. Specifically, the next cell $\mathbf{x}_{k+1}$ is determined as

\begin{equation}
    \mathbf{x}_{k+1}
    = \argmax_{\mathbf{x}_{n} \in \mathcal{N}(\mathbf{x}_{k})}
    \left(
        \|\mathbf{x}_{\mathrm{dest}}-\mathbf{x}_{k}\|_2
        -
        \|\mathbf{x}_{\mathrm{dest}}-\mathbf{x}_{n}\|_2
    \right),
    \label{eq:local_step}
\end{equation}

\noindent where $\mathbf{x}_{n}$ is the neighboring cell of $\mathbf{x}_k$. Repeating this procedure defines the spreading path $\mathcal{P}(\mathbf{x}_{\mathrm{src}},\mathbf{x}_{\mathrm{dest}})$. \\

\noindent \textbf{Updating biofilm density and mass fraction (STEP 2-4)}: Once the excess cells and spreading paths are identified, we will update the biofilm density along the determined spreading path. For consistent spread/movement, we need to conserve both the biofilm density (below the threshold $M_{\max}$) and the mass fraction of each bacterial species to capture multi-species biofilm stratification \cite{Tang2013}. Accordingly, we first update the biofilm density using \Cref{eq:spreading_mass} and then compute the species mass fractions from a mixture of the biofilms in the cell $l+1$ and the incoming biofilms transferred from the upstream cell $l$. 

Let the spreading path be indexed by $l = 0,1,\dots,L$, where $l=0$ refers to the source cell, and $l \geq 1$ refers to successive cells along the spreading path. At each iteration, excess biomass is removed from the upstream cell $l$ and transferred to the cell $l+1$. Here, $M_l$ and $M_{l+1}$ are the biofilm density before spreading, and $\widetilde{M}_l$ and $\widetilde{M}_{l+1}$ refer to the biofilm density after spreading. We compute the biofilm density after spreading and the mass fraction of each species as 

\begin{subequations}
\label{eq:spreading_mass}
\begin{gather}
    \mathcal{S}_{l} = \min\left(\frac{{S}_{0}V_0}{V_l}, M_{l} - M_{\max}\right),\\
    \widetilde{M}_{l} = {M}_{l} - \mathcal{S}_{l},\\
    \widetilde{M}_{l+1} = {M}_{l+1} + \mathcal{S}_{l}\frac{V_l}{V_{l+1}}, 
\end{gather}
\end{subequations}

and

\begin{subequations}
\label{eq:spreading_fraction}
\begin{gather}
    \widetilde{f}_{i,0} = f_{i,0},\\  
    \widetilde{f}_{i,l+1} = \frac{M_{l+1} f_{i,l+1} + S_l \dfrac{V_l}{V_{l+1}} \widetilde{f}_{i,l}}{\widetilde{M}_{l+1}},
    \end{gather}
\end{subequations}

\noindent where $\mathcal{S}_{0}$ and $\mathcal{S}_{l}$ refer to initial excess and excess at step $l$, respectively, $V_{l}$ and $V_{l+1}$ are the cell volumes at steps $l$ and $l+1$, respectively, and $\widetilde{f}_{i,l}$ is the mass fraction of bacterial species $i$ at step $l$ after spreading. To ensure mass conservation during spreading on non-uniform meshes, we incorporate the volume ratio $V_l / V_{l+1}$ into the redistribution calculation. \Cref{fig:CA_app1} STEP 2-4 shows a representative scenario for directional spreading.

\subsection{Numerical implementation of the DBS-CA algorithm}
\label{algorithm}

The computational framework for simulating the coupled dynamics of permeable biofilms and fluid flow is summarized as follows:

\begin{description}
\item[Step 1.] Solve \Cref{eq:dbs_eq1,eq:dbs_eq2} to obtain the velocity and pressure fields.

\item[Step 2.] Solve \Cref{eq:substrates} and \Cref{eq:bacteria} to update the concentration fields of substrates, suspended bacteria, and biofilms.

\item[Step 3.] Update the states of the biofilm cells using \Cref{eq:cell_state}.

\item[Step 4.] Perform stochastic spread of the excess biofilm to neighboring cells using Step 1 in \Cref{fig:CA_app1}.

\item[Step 5.] Perform directional spreading of the remaining excess biofilm along the shortest path to the interfacial cells. The density and volume fractions of each biofilm species are then updated using \Cref{eq:spreading_mass} and \Cref{eq:spreading_fraction}, respectively.

\item[Step 6.] Update porosity and permeability based on updated biofilm density using \Cref{eq:vf} and \Cref{eq:KC}, respectively.
\end{description}

\begin{figure}
  \centering
    \centerline{
        \includegraphics[scale=0.5]{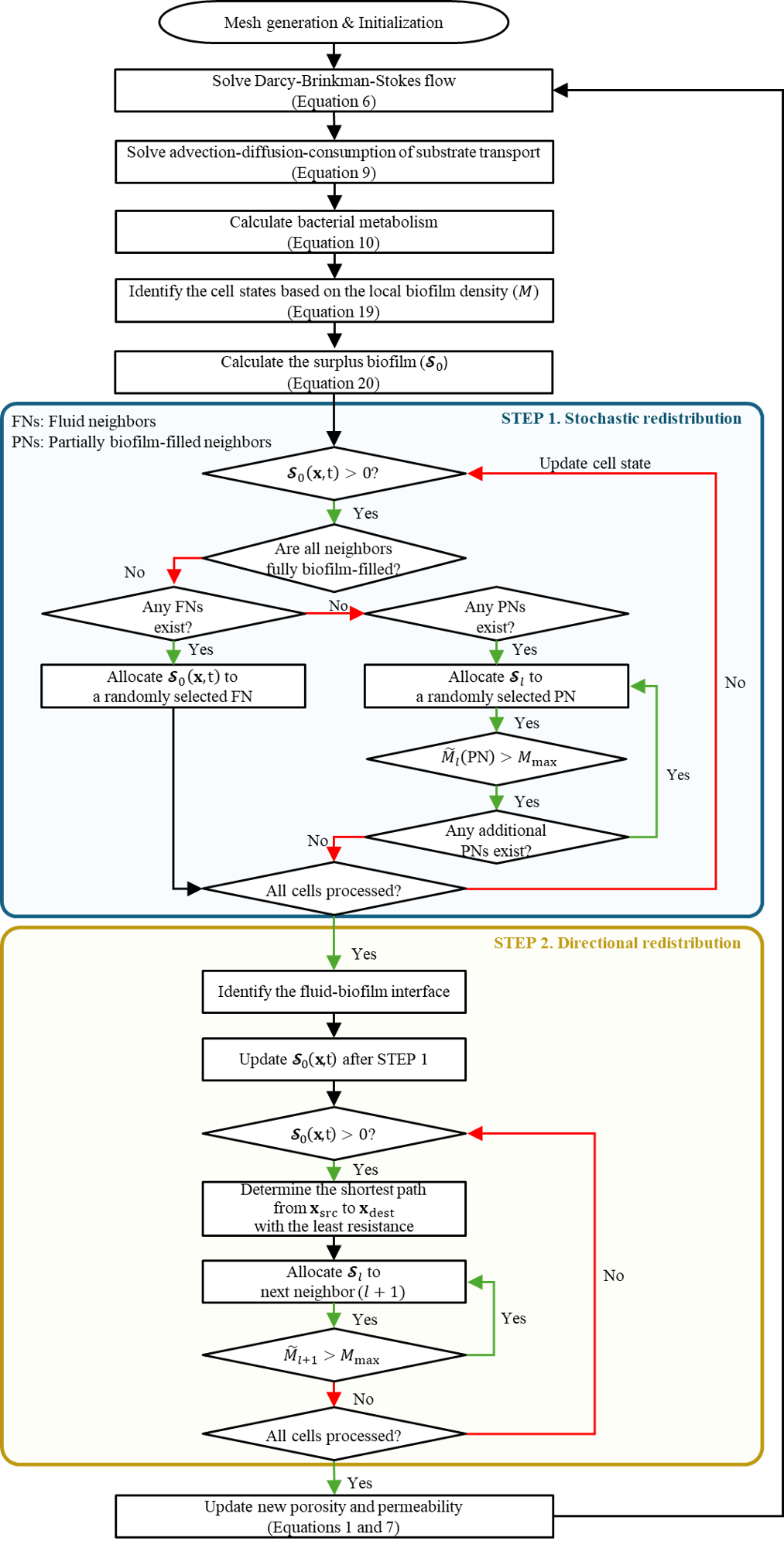}} 
    \caption{Flowchart of the proposed DBS-CA framework. The algorithm couples flow, transport, and biofilm dynamics with two-stage stochastic and directional redistribution.}
    \label{fig:CA_flow}
\end{figure}

\noindent A detailed flow chart of the algorithm is provided in \Cref{fig:CA_flow}.

\section{Numerical verification}
\label{verification}

In this section, we assess the proposed framework using three test problems. Cases 1 (\Cref{sec:finger}) and 2 (\Cref{sec:strati}) focus on whether the method qualitatively reproduces established benchmark behaviors for diffusion-dominated biofilm growth \cite{Picioreanu1998b} and multispecies stratification \cite{Tang2013}, whereas Case 3 provides a quantitative convergence study for coupled flow and biofilm dynamics in a nontrivial geometry. 

\subsection{Case 1 : Diffusion-limited fingering in a single-species biofilm}
\label{sec:finger}

Here, we consider a diffusion-only system that involves oxidizing ammonia (substrate) by biofilms consisting of nitrifiers such that

\begin{align}
\mathrm{C_1}
\xrightarrow{\text{growth}} \mathrm{X_{1}^\text{b}}.
\end{align}

\noindent For the case of a single substrate and a single microbial species in the biofilm \cite{Picioreanu1998b}, the governing equations for  substrate and biofilm are given by

\begin{subequations}
\begin{align}
\frac{\partial (\varepsilon_{\text{f}} \overline{C}_{1})}{\partial t}
- \nabla \cdot (\varepsilon_{\text{f}} D^*_{1} \nabla \overline{C}_{1})= - \varepsilon_{\text{f}}(\hat{q}^{\max}_{1}+b_1)\frac{\overline{C}_{1}}{\overline{C}_{1} + K_1}\overline{X}_{1}^\text{b},
\end{align} 

\begin{align}
\frac{\partial (\varepsilon_{\text{f}} \overline{X}_{1}^\text{b})}{\partial t}
 = \varepsilon_{\text{f}}Y_1\hat{q}^{\max}_{1}\frac{\overline{C}_{1}}{\overline{C}_{1} + K_1}\overline{X}_{1}^\text{b}
 - \varepsilon_{\text{f}} b_1 \overline{X}_{1}^\text{b}. 
\end{align}

\end{subequations}

\noindent To demonstrate the capability of the proposed method in various reaction-transport regimes, we consider the ratio of biofilm growth to substrate transport such that 

\begin{align}
G = Y_1 Da=Y_1H_b^2\frac{\hat{q}^{\max}_{1}M_{\max}}{D_1^*C_{1}^0},
\end{align}

\noindent where $Da$ is the Damkh\"{o}ler number, $H_{b}$ is the maximum thickness of the biofilm (in this study, the height of the channel), and $C_{1}^0$ is the bulk concentration of substrate.

\begin{figure}
  \centering
    \centerline{
        \includegraphics[scale=0.6]{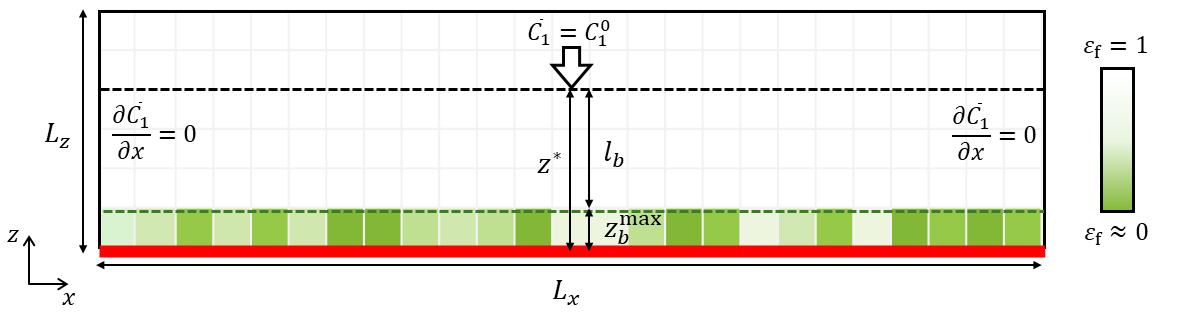}} 
    \caption{Schematic of the computational setup for Case 1, with biofilms shown in green and the solid surface in red.}
    \label{fig:setup_case1}
\end{figure}

Figure~\ref{fig:setup_case1} shows the computational domain for Case 1. For biofilm initialization, we prescribe the first layer of cells with $\varepsilon_\text{b}$ randomly varying between 0.5 and 1.0. Zero-gradient boundary conditions are imposed on all boundaries. To ensure that substrate transport does not become the limiting step, we enforce

\begin{equation}
\overline{C}_{1}(x,y,z,t) = C_1^0 \quad \text{for } z > z^*,
\end{equation}

\noindent where the reference height $z^*$ is defined as $z^* = z_b^{\max} + l_b$, $z_b^{\max}$ is the maximum biofilm height at time $t$, and $l_b = \SI{4e-5}{\meter}$ is the prescribed thickness of the diffusion boundary layer. A detailed summary of the parameters used in Case 1 can be found in \Cref{tab:ex1}.

\begin{table}[t]
\centering
\begin{threeparttable}
\caption{Model parameters used in Example~1. Parameter values are taken from~\cite{Picioreanu1998b}.}
\label{tab:ex1}
\begin{tabularx}{\textwidth}{@{}l l X@{}}
\toprule
Parameter & Unit & Value \\
\midrule
Domain size, $L_x \times L_z$ & \unit{\meter} & $0.002 \times 0.0004$ \\
Grid spacing, $\Delta x$ & \unit{\meter} & $4\times 10^{-6}$ \\
Substrate diffusion coefficient, $D_1^*$ & \unit{\square\metre\per\second} & \num{1.6e-9} \\
Bulk substrate concentration, $C_1^0$ & \unit{\kilogram\per\cubic\metre} & 
\SIlist{21.3e-2;10.6e-2;2.13e-2;0.43e-2}{} \\
Maximum biofilm density, $M_{\max}$ & \unit{\kilogram\per\cubic\metre} & \SI{70}{} \\
Dry biomass density, $\rho_\text{b}^0$ & \unit{\kilogram\per\cubic\metre} & \SI{350}{} \\
Half-saturation constant, $K_1$ & \unit{\kilogram\per\cubic\metre} & \SI{3.5e-4}{} \\
Yield coefficient, $Y_1$ & -- & 0.045 \\
Maintenance coefficient, $b_1$ & \unit{\per\second} & \SI{3e-5}{} \\
Maximum substrate utilization rate, $\hat{q}^{\max}_{i}$ & \unit{\per\second} & \SI{3.38e-4}{} \\
Biofilm intrinsic permeability, $k_0$ & \unit{\square\metre} & \num{1e-9}  \\
\bottomrule
\end{tabularx}
\end{threeparttable}
\end{table}

\begin{figure}
  \centering
    \centerline{
        \includegraphics[scale=0.5]{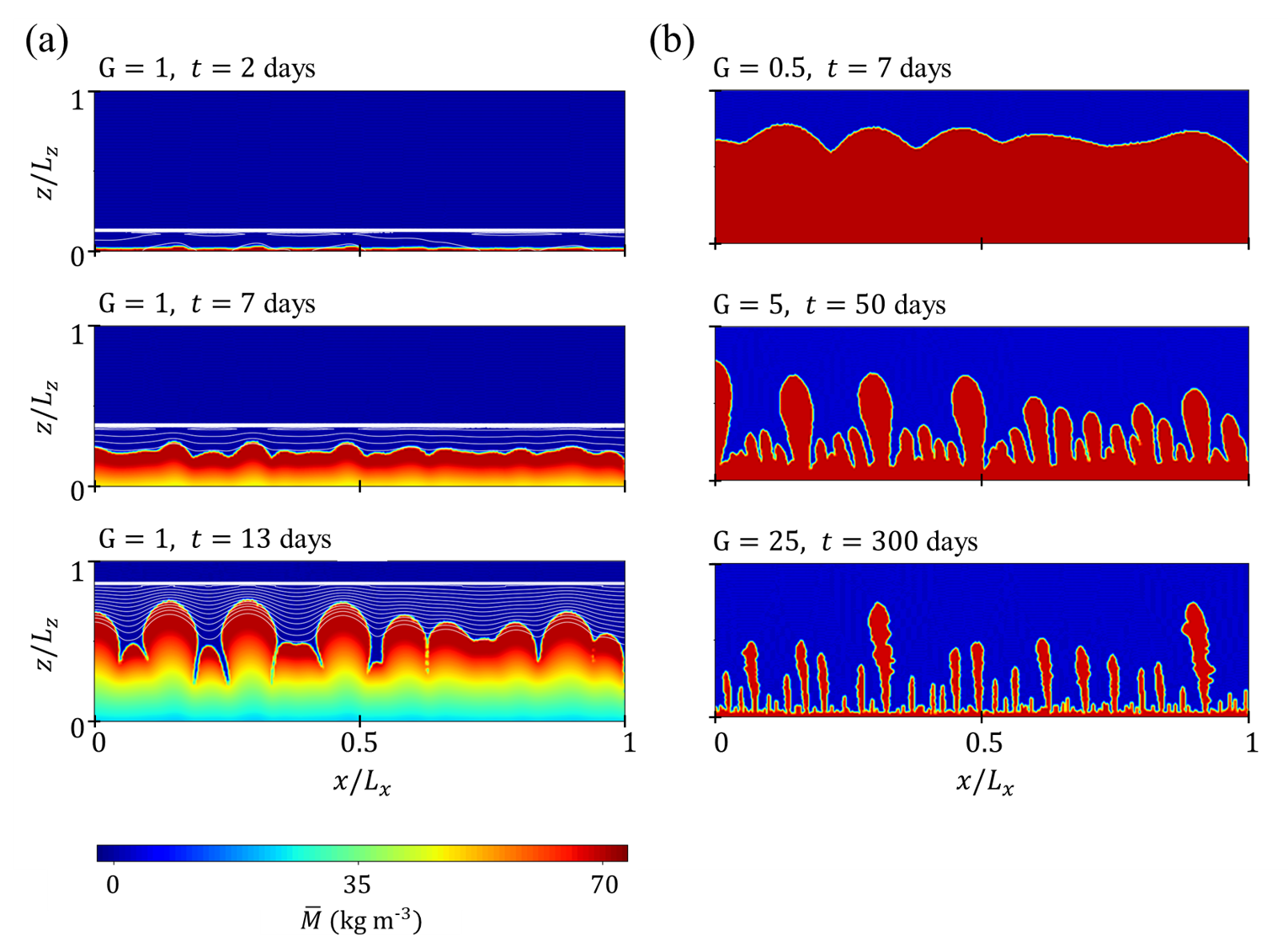}} 
    \caption{Biofilm development at different values of $G$. (a) Temporal evolution for $G=1$ at $t=2$, 7, and 13 days (top to bottom), with contour lines indicating substrate $C_1$ concentrations at 5\% increments. (b) Binary biofilm field at the time when $z_b^{\max} = 0.3~\si{\milli\meter}$ for $G=0.5$, 5, and 25 (top to bottom).}
    \label{fig:finger}
\end{figure}

Figure~\ref{fig:finger}a shows the temporal evolution of the biofilm and the associated substrate. Initially, a nearly uniform biofilm layer covers the solid surface ($t=2~\text{days}$). As the substrate becomes limited, biofilms with greater access to the substrate continue to grow, whereas those with limited access grow slower ($t=7~\text{days}$), leading to the formation of fingering structures ($t=13~\text{days}$). The predicted biofilm structures are qualitatively consistent with previous studies~\cite{Picioreanu1998b, Alpkvista2007, Benioug2017, Zang2025}.

To evaluate the effects of $G$ on biofilm fingering, we simulate four cases in which $G$ is varied between 0.5 and 25 by changing the bulk substrate concentration ($C_j^0$, \Cref{tab:ex1}). All simulations are performed until the thickness of the biofilm reaches $0.75L_z$. Figure~\ref{fig:finger}b shows that small $G$ yields fewer but larger fingers, whereas large $G$ yields more but smaller fingers. This trend is consistent with the expected transition from growth-limited to transport-limited conditions and supports the ability of the proposed method to reproduce the qualitative dependence of the morphology of the biofilms on $G$.

\subsection{Case 2: Competition-driven stratification in a multi-species biofilm} 
\label{sec:strati}

Case 2 focuses on stratification of multispecies biofilms competing for multiple substrates in a diffusion-dominant system. We denote ${C}_1$, ${C}_2$, and ${C}_3$ as concentrations of acetate, ammonium, and dissolved oxygen, respectively. For biofilms, we consider three different species which are heterotrophs ($X_{1}^\text{b}$), autotrophs ($X_{2}^\text{b}$) and inert biomass ($X_{3}^\text{b}$). The biological reactions are given by

\begin{subequations}
\begin{align}
\mathrm{C_1 + C_3}
\xrightarrow{\text{growth}} \mathrm{X_{1}^\text{b}} \xrightarrow{\text{decay}} \mathrm{X_{3}^\text{b}} ,
\end{align}
\begin{align}
\mathrm{C_2 + C_3}
\xrightarrow{\text{growth}} \mathrm{X_{2}^\text{b}} \xrightarrow{\text{decay}} \mathrm{X_{3}^\text{b}}.
\end{align}
\end{subequations}

\begin{figure}
  \centering
    \centerline{
        \includegraphics[scale=0.5]{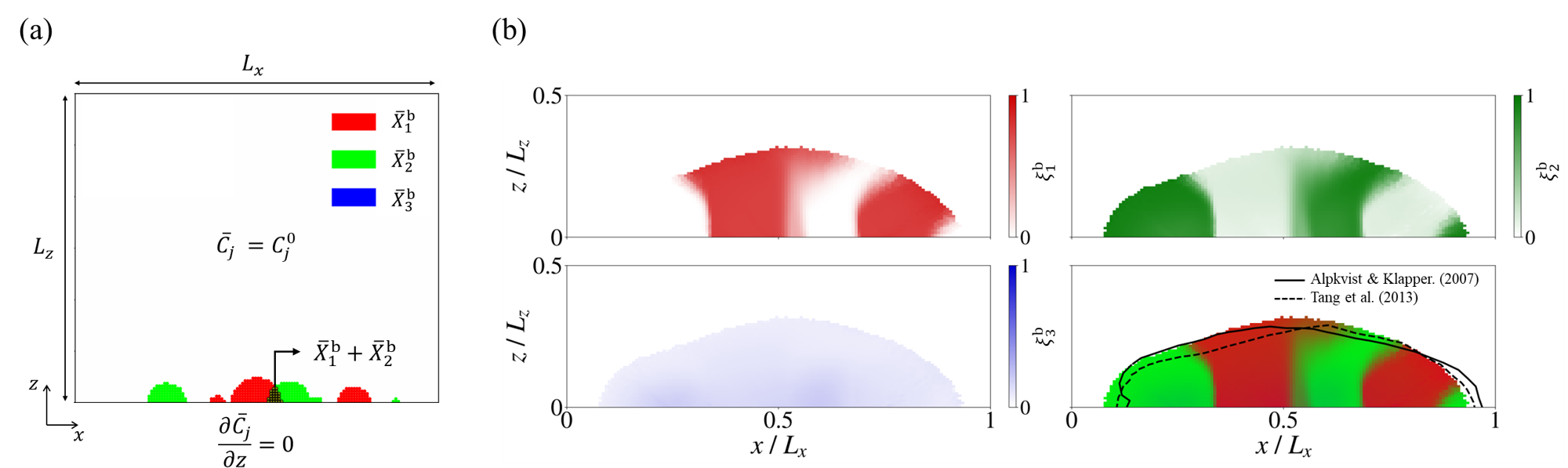}} 
    \caption{Multispecies biofilm stratification (Case 2). (a) Computational setup at $t=0$, showing the initial distribution of the attached species and the substrate boundary conditions. (b) Biofilm structure at $t=3.5$ days. The top-left, top-right, and bottom-left panels show the mass fractions of biofilm species $1$, $2$, and $3$, respectively, defined as $\xi_i^b=\varepsilon_f \overline{X}_i^b/\overline{M}$. The bottom-right panel shows the total biofilm density $\overline{M}$, with reference contours from \cite{Alpkvista2007, Tang2013}.}
    \label{fig:case2}
\end{figure}

\noindent The corresponding governing equations for substrates and biofilms are given in \ref{app:case2}.

Figure~\ref{fig:case2}a shows the schematic of the computational domain for Case 2. The initial biofilms, which consist of multiple colonies of $\overline{X}_{1}^\text{b}$ (red) and $\overline{X}_{2}^\text{b}$ (green), are placed on the bottom surface. For $x = 0.367$-$0.4~\unit{\milli\meter}$, we initialize a mixed colony with $\overline{X}_{1}^\text{b}$ and $\overline{X}_{2}^\text{b}$ with equal volume fractions. Following \cite{Alpkvista2007, Tang2013}, we impose a zero-gradient boundary condition at the bottom boundary ($z=0$) for all species. To ensure that biofilm growth is not limited by substrate transport, we enforce $\overline{C}_i=C_i^0$ for all locations with $\epsilon_{\text{f}} = 1$. A detailed summary of the parameters used in Case 2 is shown in \Cref{tab:ex2}. Figure \ref{fig:case2}b compares the mass fraction of each species and total biofilm density using the new CA method with those obtained from \cite{Alpkvista2007, Tang2013} at time $t=3.5~\text{days}$. Our results indicate that the improved CA method reproduces multispecies biofilm patterns that are qualitatively consistent with those reported in \cite{Alpkvista2007, Tang2013}. Specifically, the improved CA method captures the multispecies biofilm stratification and mixed transition region, supporting its ability to preserve multispecies structure during redistribution.

\begin{table}[t]
\centering
\begin{threeparttable}
\caption{Model parameters used in Example~2. Parameter values are taken from~\cite{Alpkvista2007, Tang2013}.}
\label{tab:ex2}
\begin{tabularx}{\textwidth}{@{}l l X@{}}
\toprule
Parameter & Unit & Value \\
\midrule
Domain size, $L_x \times L_z$ & \unit{\meter} & $0.0007 \times 0.0006$ \\
Grid spacing, $\Delta x$ & \unit{\meter} & $5\times 10^{-6}$  \\
Substrate diffusion coefficient, $D_i^*$ & \unit{\square\metre\per\second} & $D_1^*=\num{9.6e-10}$, $D_2^*=\num{1.7e-9}$, $D_3^*=\num{2e-9}$ \\
Bulk substrate concentration, $C_i^0$ & \unit{\kilogram\per\cubic\metre} & $C_1^0=\num{3e-3}$, $C_2^0=\num{13e-3}$, $C_3^0=\num{8e-3}$ \\
Maximum biofilm density, $M_{\max}$ & \unit{\kilogram\per\cubic\metre} & \num{5} \\
Dry biomass density, $\rho_\text{b}^0$ & \unit{\kilogram\per\cubic\metre} & \num{25} \\
Half-saturation constant, $K_j$ & \unit{\kilogram\per\cubic\metre} & $K_1=\num{5e-3}$, $K_2=\num{1e-3}$, $K_{31}=\num{5e-3}$, $K_{32}=\num{1e-4}$ \\
Yield coefficient, $Y_j$ & -- & $Y_1=\num{4e-1}$, $Y_2=\num{22e-2}$ \\
Endogenous decay rate constant, $b_i$ & \unit{\per\second} & $b_1=\num{2.3e-6}$, $b_2=\num{5.8e-7}$ \\
Inactivation rate constant, $p_i$ & \unit{\per\second} & $p_1=\num{1.2e-6}$, $p_2=\num{1.2e-6}$ \\
Conversion factor, $\alpha_j$ & -- & $\alpha_1=\num{1}$, $\alpha_2=\num{4.6}$ \\
Maximum substrate utilization rate, $\hat{q}^{\max}_{i}$ & \unit{\per\second} & $\hat{q}^{\max}_{1}=\num{1.4e-4}$, $\hat{q}^{\max}_{2}=\num{5e-5}$ \\
Biofilm intrinsic permeability, $k_0$ & \unit{\square\metre} & \num{1e-9}  \\
\bottomrule
\end{tabularx}
\end{threeparttable}
\end{table}

\subsection{Case 3: Grid convergence study under flow conditions}

In this section, we focus on demonstrating the capability of the developed method to simulate biofilm dynamics under complex flow. To assess the numerical accuracy and spatial convergence of the proposed framework, we perform a grid-convergence study on a rectangular-bump geometry under uniform flow. The system consists of a single microbial species and a single substrate, with growth, decay, attachment, and detachment processes taken into account. The biological processes are given by

\begin{equation}
    \text{C}_1 \xrightarrow{\text{growth}} X_1^{\text{b}} \xrightarrow{\text{detachment}} X_1^{\text{sb}},
\end{equation}

\noindent where the superscripts $\mathrm{b}$ and $\mathrm{sb}$ refer to biofilms and suspended bacteria, respectively. The governing equations for substrate transport and microbial dynamics are given in \ref{Eqs_Sec3}. 

\begin{figure}
  \centering
    \centerline{
        \includegraphics[scale=0.5]{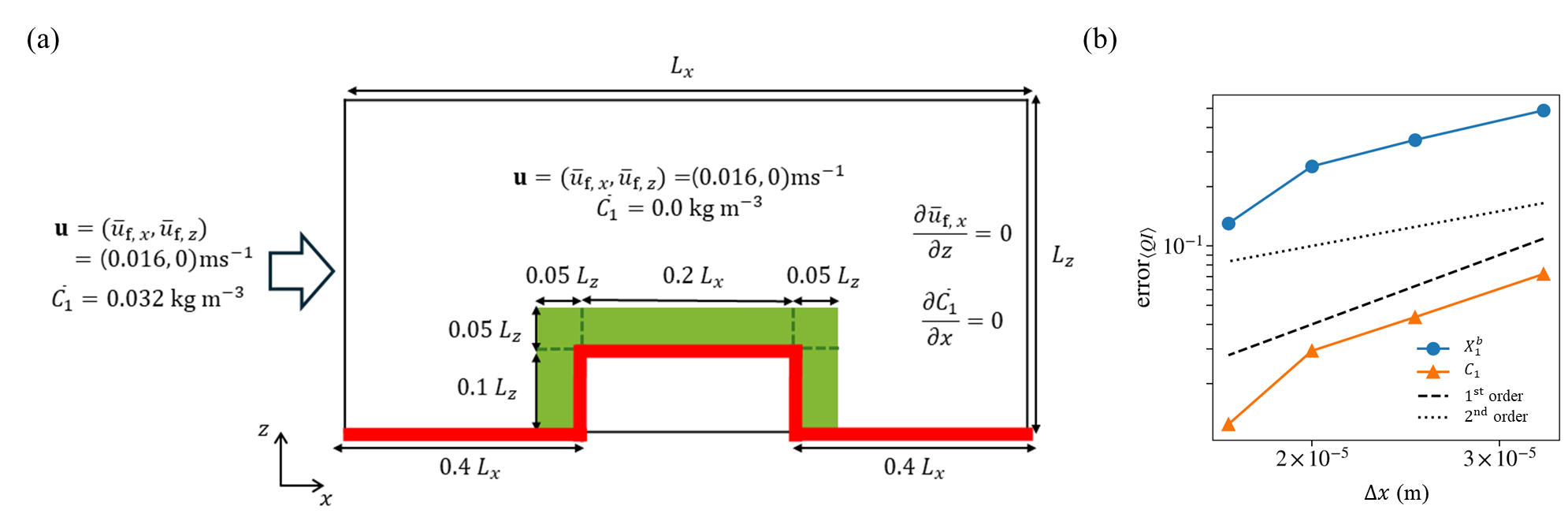}} 
    \caption{Grid convergence study for biofilms and substrates under flow over a rectangular bump (Case~3). (a) Schematic of the computational setup with initial biofilm distribution in green and the solid surface in red. (b) Grid convergence results for biofilm and substrate concentrations.}
    \label{fig:case3_grid}
\end{figure}

\Cref{fig:case3_grid}(a) shows the schematic of the computational domain, where $L_x = \SI{0.003}{\meter}$ and $L_z = \SI{0.002}{\meter}$ are the length and height of the domain, respectively. A rectangular bump with dimensions of $\SI{0.6}{\milli\meter} \times \SI{0.2}{\milli\meter}$ is placed at the center of the computational domain. At the inlet, we impose Dirichlet boundary conditions for both velocity and concentration, with $\mathbf{\overline{u}_{\text{f}}} = (u_{f,x}, u_{f,y}) = (0.016, 0)$ \unit{\meter\per\second} and $\overline{C}_1 = \SI{3.2e-4}{\kilogram\per\cubic\metre}$, respectively, and apply a zero-gradient condition for pressure. This yields a Reynolds number of $\mathrm{Re} = \dfrac{u_{f,x} L_z}{\nu_f} = 32$, where $\nu_f$ is the kinematic viscosity of water. At the outlet, we prescribe zero-gradient boundary conditions for both velocity and concentration and impose a Dirichlet condition of $p = 0$ for the pressure. At the top and bottom boundaries, we apply zero-gradient conditions to both concentration and pressure and enforce no-slip boundary conditions for the velocity. The microbial parameters are identical to those given in \Cref{tab:ex1}. The threshold shear stress $\tau_{\max}$ for biofilm detachment and the rate of detachment $k_{\text{det}}$ are \SI{0.08}{Pa} and \SI{0.1}{\per\second}, respectively. 

For initialization, we prescribe a single-species biofilm layer of thickness $\SI{1}{\milli\meter}$ around the rectangular bump, with an initial biofilm density equal to $M_{\max}$. The velocity field is initialized as $\mathbf{\overline{u}_{\text{f}}} = (u_{f,x}, u_{f,y}) = (0.016, 0)$, and the concentration is initially zero. To perform the convergence study, we successively refine the mesh with grid spacings $\Delta x =  3.3\times 10^{-5}, \ 2.5\times 10^{-5}, \ 2\times 10^{-5},\  1.67\times 10^{-5},$ and $1.435 \times 10^{-5}\,\si{\meter}$. To maintain a constant Courant number, the timestep size $\Delta t$ is refined proportionally to the grid spacing. Simulations are performed until $t = 10$ \unit{days}. Here, we define the error as 


\begin{equation}
\mathrm{err}_{QI}
=
\frac{
\left\|
 \overline{\langle QI \rangle}_{\Delta x}(x,z)
- \overline{\langle QI \rangle}_{\Delta x_{\text{ref}}}(x,z)
\right\|_{\textit{fro}}
}{
\frac{1}{N_x N_z}
\sum_{i=1, j=1}^{N_x N_z}
\overline{\langle QI \rangle}_{\Delta x_{\text{ref}}}(x_i,z_j)
}
\end{equation}

\noindent where $\Delta x_{\text{ref}}=1.435 \times 10^{-5}$ \unit{\meter}, $QI$ is the quantity of interest, and $\| \cdot \|_\textit{fro}$ is the Frobenius norm operator. 

Figure~\ref{fig:case3_grid}(b) shows $\text{err}_{\textit{$X_1^b$}}$ and $\text{err}_{\textit{$C_1$}}$ as functions of grid spacing $\Delta x$. In general, we observe second-order convergence with respect to $\Delta x$ for $\overline{X}_1^b$ and $\overline{C}_1$, demonstrating the accuracy of the proposed method. This result confirms that the proposed framework retains numerical consistency in the presence of evolving biofilm interfaces under uniform flow conditions.

\section{Application to stratification in counter-diffusional biofilms}
\label{sec:counter-biofilms}

Counter-diffusional biofilms are commonly found in engineered systems \cite{Nerenberg2016-qx, Martin2015-vn, Shukla2026-is}, where stratified biofilms develop under varying flow conditions. In such systems, the dynamics are governed by substrates entering from opposite boundaries of the biofilms. In contrast to co-diffusive biofilms (Cases 1-3), in which all substrates enter from the same side, counter-diffusion generally leads to spatially separated reaction zones, evolving active layers, and biofilm stratification. These features make such systems a suitable test case for assessing whether the proposed framework can consistently couple substrate transport, multispecies kinetics, and biofilm spreading.

In this section, we consider a membrane-aerated biofilm reactor \cite{Shanahan2004} as a canonical example of a counter-diffusional system to demonstrate the capability of the proposed framework in simulating strongly heterogeneous and dynamically evolving multispecies biofilms. In this configuration, dissolved oxygen ($C_3$) is supplied from the solid surface, whereas organic carbon ($C_4$) and ammonium ($C_1$) are delivered from the bulk liquid. Within the biofilm, nitrate ($C_2$) is produced as an intermediate species. Two microbial species are considered, nitrifiers and denitrifiers. Species-dependent adhesion is not considered in this application. The resulting stratification therefore arises from counter-diffusional substrate supply, microbial kinetics, and biomass redistribution. Nitrifiers ($X_{1}^\text{sb}$ or $X_{1}^\text{b}$) consume $C_1$ and $C_3$ and produce $C_2$, whereas denitrifiers ($X_{2}^\text{sb}$ or $X_{2}^\text{b}$) consume $C_2$ and $C_4$. The biological reactions are 

\begin{subequations}
\begin{align}
\mathrm{C_1 + C_3}
\xrightarrow{\text{growth}} \mathrm{X_{1}^\text{sb}} \text{ or } \mathrm{X_{1}^\text{b}}\xrightarrow{\text{product}} \mathrm{C_{2}} ,
\end{align}
\begin{align}
\mathrm{C_2 + C_4}
\xrightarrow{\text{growth}} \mathrm{X_{2}^\text{sb}} \text{ or } \mathrm{X_{2}^\text{b}} \xleftarrow{\text{inhibition}} \mathrm{C_{3}}.
\end{align}
\end{subequations}

\noindent The simplified governing equations for substrate transport and microbial dynamics are given in \ref{sec:counter}.

\begin{figure}
  \centering
    \centerline{
        \includegraphics[scale=0.7]{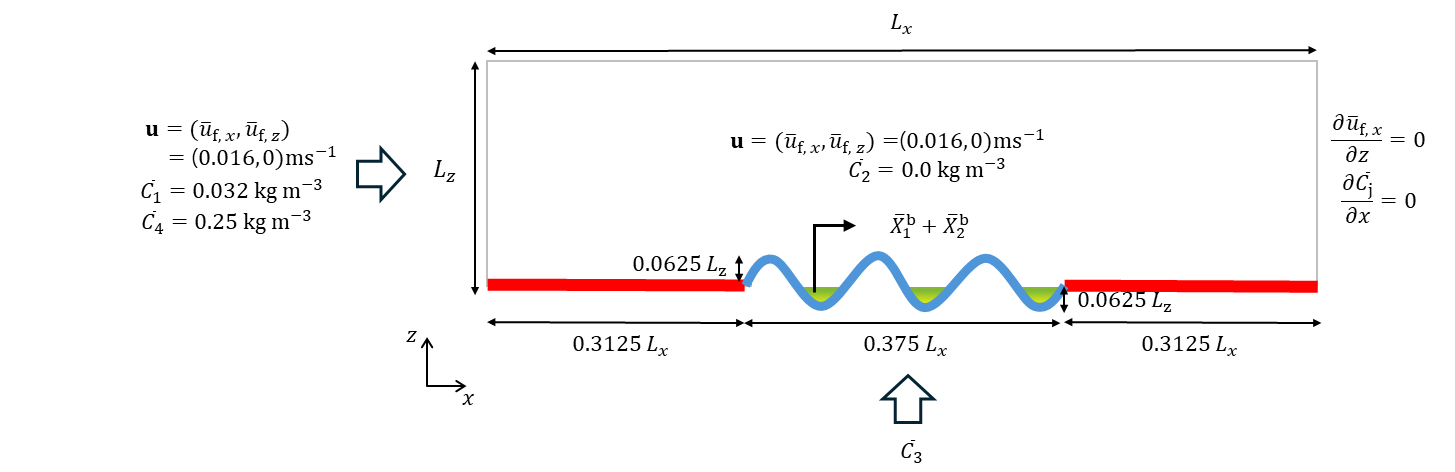}} 
    \caption{Schematic of the computational setup. The initial biofilm distribution (green) is prescribed with equal fractions of $\overline{X}_1^b$ and $\overline{X}_2^b$. The solid wall and membrane are shown in red and blue, respectively.}
    \label{fig:case4}
\end{figure}

\Cref{fig:case4} shows the schematic of the computational domain for counter-diffusional biofilms. The bottom boundary consists of two regions, a wall (red) and a membrane (blue). For velocity, we impose a Dirichlet boundary condition with a uniform inflow velocity $\overline{\mathbf{u}}_\text{f} = (0.016, 0)$ at the inlet, apply a zero-gradient condition at the outlet, and enforce no-slip conditions at the top and bottom (wall and membrane) boundaries. For pressure, a fixed value of $\overline{p}_\text{f} = 0$ is imposed at the outlet, while zero-gradient conditions are applied elsewhere. For the substrates, fixed concentrations are prescribed at the inlet, and zero-gradient conditions are applied at the top boundary and the outlet. At the bottom wall, zero-gradient conditions are imposed for all substrates. At the membrane, a Robin boundary condition is imposed for $C_3$, with the boundary concentration determined from partial pressure of oxygen using Henry’s law, such that

\begin{equation}
k_{\text{mem}} C_3+D^*_3\frac{\partial C_3}{\partial \mathbf{n}}= k_{\text{mem}} K_HP_{C_3},
\end{equation} 

\noindent where $k_{\text{mem}}$ is the membrane gas permeability, $K_H$ is Henry’s constant, $\mathbf{n}$ is the normal vector pointing away from membrane, and $P_{C_3}$ is the partial pressure of oxygen. For all other substrates, zero-gradient conditions are applied at the membrane. 

At $t = 0$, all substrate concentrations and pressure are initialized to zero, and the velocity field is initialized as $\overline{\mathbf{u}}_\text{f} = (u_{f,x}, u_{f,y}) = (0.016, 0)$. For biofilms, equal initial concentrations of $\overline{X}_1^b = \overline{X}_2^b = 5~\unit{\kilogram\per\cubic\metre}$ are prescribed for $-0.0625L_z \le z \le 0$. A detailed summary of the parameters used is provided in \Cref{tab:ex4}.

\begin{figure}
  \centering
    \centerline{
        \includegraphics[scale=0.5]{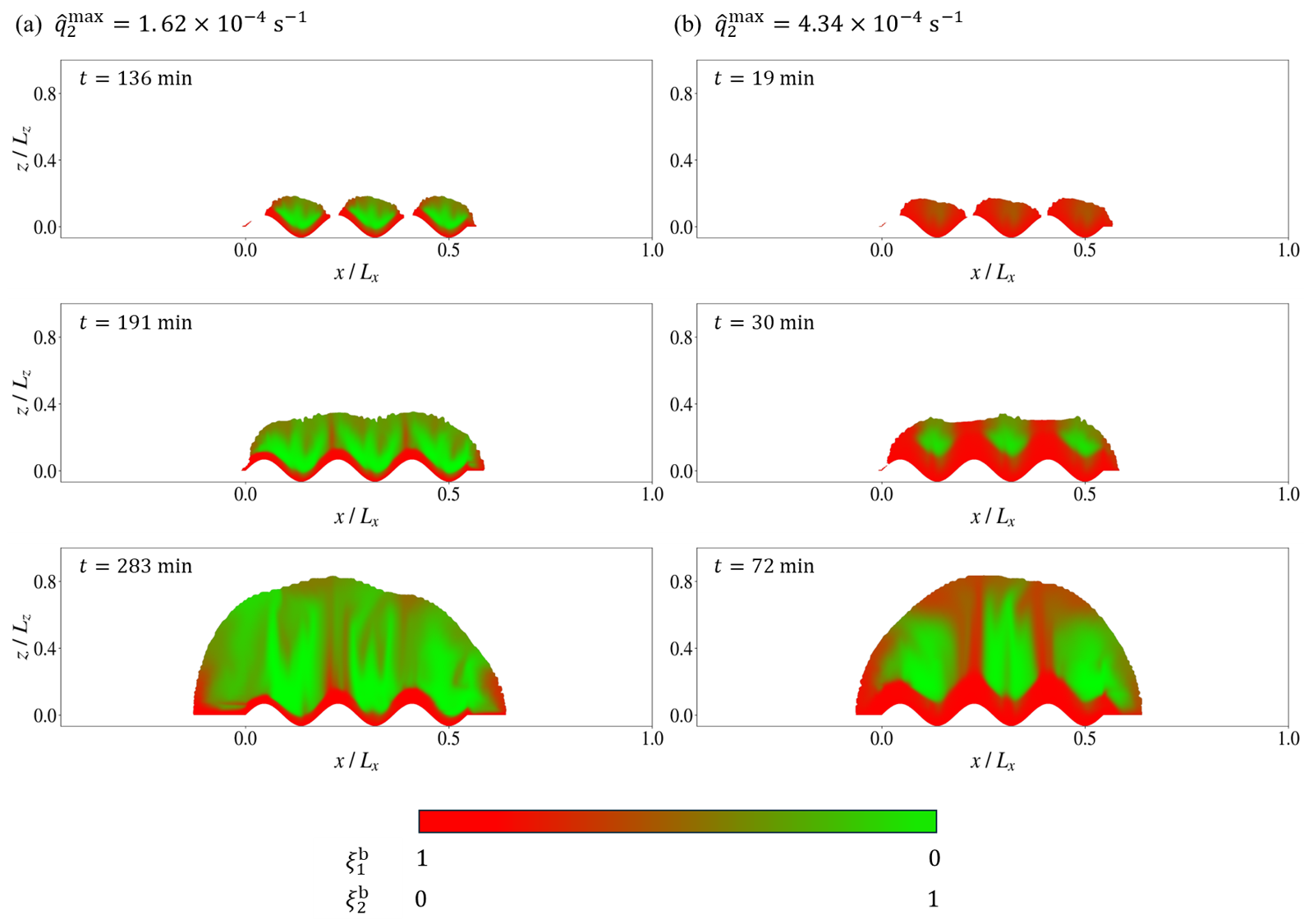}} 
    \caption{Spatiotemporal evolution of biofilms under two substrate utilization rates (left: $\hat{q}_2^{\max}=1.62\times10^{-4}\si{\per\second}$; right: $\hat{q}_2^{\max}=4.34\times10^{-4}\si{\per\second}$). Rows show successive time instances from early to late growth.}
    \label{fig:MABR1}
\end{figure}

Figure~\ref{fig:MABR1} shows the temporal evolution of biofilms for different values of $\hat{q}^{\max}_{2}$. In both cases, we observed a stratified biofilm structure, with $X_1^b$ concentrated near the membrane and $X_2^b$ occupying the outer biofilm region. As $\hat{q}^{\max}_{2}$ increases,  the spatial distribution of the two species shifts, leading to a thicker region occupied by $X_1^b$. To quantify the biofilm structure and the associated substrate fields, we compute the $x$-directional average of a generic quantity $QI$ over the biofilm-occupied region as
\begin{equation}
    \langle QI \rangle_x (z,t)
    =
    \frac{\displaystyle \sum_{i=1}^{N_x} QI_i(z,t)\, \mathcal{I}_i(z,t)}
    {\displaystyle \max\!\left(\sum_{i=1}^{N_x} \mathcal{I}_i(z,t),\ \epsilon \right)},
    \label{eq:xavg_biofilm_discrete}
\end{equation}

\noindent where $QI_i$ refers to the value of the quantity of interest (e.g., $X_i^{b}$, $C_i$, etc.) at grid point $i$, $N_x$ is the number of grid cells in the $x$ direction, and the indicator function $\mathcal{I}_i$ is defined as
\begin{equation}
    \mathcal{I}_i(z,t)
    =
    \begin{cases}
        1, & \varepsilon_{f,i}(z,t) < 1, \\
        0, & \text{otherwise},
    \end{cases}
\end{equation}

\noindent where $\epsilon$ is a small positive threshold for preventing division by zero.

\begin{figure}
  \centering
    \centerline{
        \includegraphics[scale=0.45]{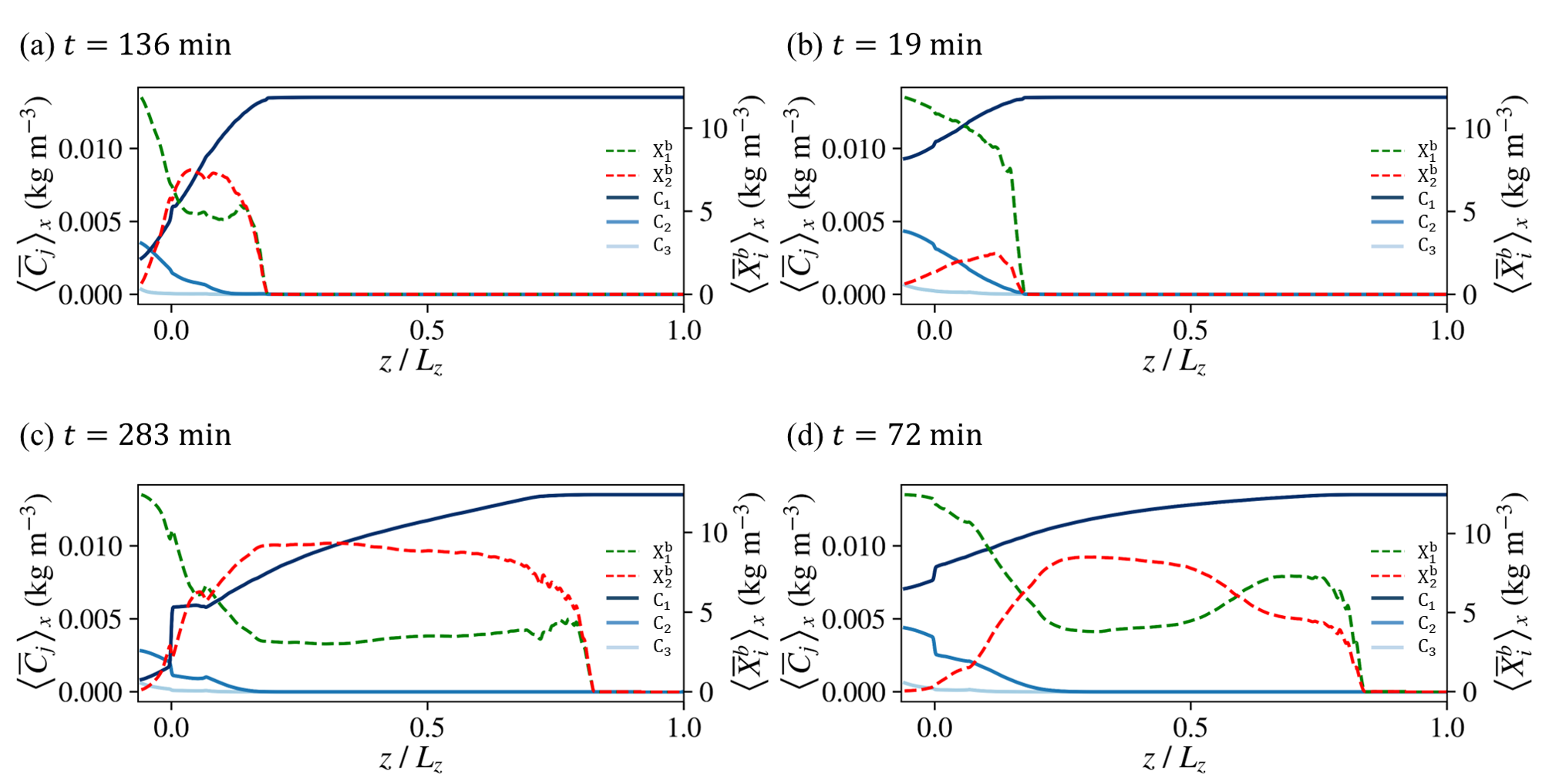}} 
    \caption{Average concentrations of biofilm species (left axis) and substrates (right axis) as functions of $z$ for $\hat{q}_2^{\max}=1.62\times10^{-4}~\si{\per\second}$ at (a) $t=136$ min and (b) $t=283$ min, and for $\hat{q}_2^{\max}=4.34\times10^{-4}~\si{\per\second}$ at (c) $t=19$ min and (d) $t=72$ min.}
    \label{fig:MABR2}
\end{figure}

Figure \ref{fig:MABR2} shows the normalized streamwise average concentrations of biofilm species and substrates as functions of $z$. We observe that $X_1^b$ dominates in regions of high $C_3$, while $X_2^b$ is distributed farther from the membrane, where oxygen availability decreases and exposure to oppositely supplied substrates increases. The opposing substrate gradients result in spatially separated reaction zones across the biofilm thickness. This spatial organization is characteristic of counter-diffusional biofilms and demonstrates that the proposed framework can resolve the coupled interactions among substrate transport, multispecies reactions, and biomass redistribution while maintaining distinct reaction zones across the biofilm thickness.

\begin{table}[t]
\centering
\begin{threeparttable}
\caption{Model parameters used in the MABR system. Parameter values are taken from~\cite{Shanahan2004}.}
\label{tab:ex4}
\begin{tabularx}{\textwidth}{@{}l l X@{}}
\toprule
Parameter & Unit & Value \\
\midrule
Domain size, $L_x \times L_z$ & \unit{\meter} & $0.008 \times 0.0015$ \\
Substrate diffusion coefficient, $D_i^*$ & \unit{\square\metre\per\second} & $D_1^*=\num{1.713e-9}$, $D_2^*=\num{1.655e-9}$, $D_3^*=\num{1.841e-9}$, $D_4^*=\num{9.529e-10}$ \\
Bulk substrate concentration, $C_i^0$ & \unit{\kilogram\per\cubic\metre} & $C_1^0=\num{13.5e-3}$, $C_2^0=\num{0}$, $C_4^0=\num{25e-2}$ \\
Oxygen partial pressure, $P_{C_3}$ & \unit{atm} & \num{0.021} \\
Henry's constant, $K_H$& \unit{\kilogram\per\cubic\metre} & \num{0.0448} \\
Membrane mass transfer coefficient & \unit{\metre\per\second} & \num{1e-4} \\
Maximum biofilm density, $M_{\max}$ & \unit{\kilogram\per\cubic\metre} & \num{10} \\
Dry biomass density, $\rho_\text{b}^0$ & \unit{\kilogram\per\cubic\metre} & \num{50} \\
Half-saturation constant, $K_j$ & \unit{\kilogram\per\cubic\metre} & $K_1=\num{1.4e-3}$, $K_2=\num{1e-4}$, $K_{3}=\num{4.5e-4}$, $K_{4}=\num{1.37e-2}$\\
Yield coefficient, $Y_j$ & -- & $Y_1=\num{0.2}$, $Y_2=\num{0.63}$, $Y_3=\num{0.6}$ \\
Inhibition coefficient for denitrifiers, $K_{32,\text{in}}$ & \unit{\kilogram\per\cubic\metre} & \num{1.5e-4} \\
Endogenous decay rate constant, $b_i$ & \unit{\per\second} & $b_1=\num{5.79e-7}$, $b_2=\num{4.63e-7}$ \\
Maximum substrate utilization rate, $\hat{q}^{\max}_{i}$ & \unit{\per\second} & $\hat{q}^{\max}_{1}=\num{5.79e-4}$, $\hat{q}^{\max}_{2}=\num{1.62e-4},\,\num{4.34e-4}$ \\
Biofilm intrinsic permeability, $k_0$ & \unit{\square\metre} & \num{1e-13}  \\
\bottomrule
\end{tabularx}
\end{threeparttable}
\end{table}

\section{Conclusions}

In this work, we developed a micro-continuum-CA framework for simulating multispecies biofilm dynamics in complex flows. The method couples the DBS equations, reactive transport of substrates and suspended bacteria, and biofilm dynamics with a two-stage CA algorithm for biofilm spread and interface evolution. By treating biofilms as evolving porous media, the framework provides a consistent description of fluid flow and substrate transport in both free-flow and partially occupied biofilm regions. In addition, the two-stage CA approach preserves biofilm mass fractions and concentrations on non-uniform meshes.

The developed framework was assessed using benchmark problems for diffusion-limited fingering in single-species biofilms, competition-driven stratification in multispecies biofilms, and mesh convergence under flow in a channel with a rectangular bump. The results showed that the framework reproduces established qualitative biofilm morphologies and stratification patterns. Second-order convergence was observed in the presence of flow, biofilm growth, and decay in a complex geometry.

The framework was further applied to counter-diffusional biofilms in a membrane-aerated biofilm reactor as a canonical example. The developed method captures biofilm stratification due to spatially heterogeneous reactive zones. This further demonstrates the ability of the framework to resolve coupled transport, reaction, and biomass redistribution in spatially heterogeneous multispecies systems. Overall, the proposed DBS-CA method provides a flexible approach for studying biofilm evolution in systems where interactions between flow, transport, and multispecies microbial processes are important. Further improvements, such as sharper interface representation, three-dimensional geometries, and biofilm mechanical deformation, are subjects of future research.

\appendix
\section{Governing equations for Case 2}
\label{app:case2}
\begin{subequations}
\begin{align}
\frac{\partial (\varepsilon_{\text{f}} \overline{X}_{1}^\text{b})}{\partial t}
 = \varepsilon_{\text{f}}Y_1\hat{q}^{\max}_{1}\frac{\overline{C}_{1}}{\overline{C}_{1} + K_1}\frac{\overline{C}_{3}}{\overline{C}_{3} + K_{31}}\overline{X}_{1}^\text{b}
 -\varepsilon_{\text{f}}b_1\frac{\overline{C}_{3}}{\overline{C}_{3} + K_{31}}\overline{X}_{1}^\text{b}
 -\varepsilon_{\text{f}}p_1\overline{X}_{1}^\text{b},
\end{align}
\begin{align}
\frac{\partial (\varepsilon_{\text{f}} \overline{X}_{2}^\text{b})}{\partial t}
 = \varepsilon_{\text{f}}Y_1\hat{q}^{\max}_{2}\frac{\overline{C}_{2}}{\overline{C}_{2} + K_2}\frac{\overline{C}_{3}}{\overline{C}_{3} + K_{32}}\overline{X}_{2}^\text{b}
 -\varepsilon_{\text{f}}b_2\frac{\overline{C}_{3}}{\overline{C}_{3} + K_{32}}\overline{X}_{2}^\text{b}
 -\varepsilon_{\text{f}}p_2\overline{X}_{2}^\text{b},
\end{align}
\begin{align}
 \frac{\partial (\varepsilon_{\text{f}} \overline{X}_{3}^\text{b})}{\partial t}
 = \varepsilon_{\text{f}}p_1\overline{X}_{1}^\text{b}
 +\varepsilon_{\text{f}}p_2\overline{X}_{2}^\text{b},
\end{align}
\begin{align}
\frac{\partial (\varepsilon_{\text{f}} \overline{C}_{1})}{\partial t}
- \nabla \cdot (\varepsilon_{\text{f}} D^*_{1} \nabla \overline{C}_{1})= - \varepsilon_{\text{f}}\hat{q}^{\max}_{1}\frac{\overline{C}_{1}}{\overline{C}_{1} + K_1}\frac{\overline{C}_{3}}{\overline{C}_{3} + K_{31}}\overline{X}_{1}^\text{b},
\end{align} 
\begin{align}
\frac{\partial (\varepsilon_{\text{f}} \overline{C}_{2})}{\partial t}
- \nabla \cdot (\varepsilon_{\text{f}} D^*_{2} \nabla \overline{C}_{2})= - \varepsilon_{\text{f}}\hat{q}^{\max}_{1}\frac{\overline{C}_{2}}{\overline{C}_{2} + K_2}\frac{\overline{C}_{3}}{\overline{C}_{3} + K_{32}}\overline{X}_{1}^\text{b},
\end{align} 
\begin{align}
\frac{\partial (\varepsilon_{\text{f}} \overline{C}_{3})}{\partial t}
- \nabla \cdot (\varepsilon_{\text{f}} D^*_{3} \nabla \overline{C}_{3})  &= - \varepsilon_{\text{f}}\hat{q}^{\max}_{1}\frac{\overline{C}_{1}}{\overline{C}_{1} + K_1}\frac{\overline{C}_{3}}{\overline{C}_{3} + K_{31}}\overline{X}_{1}^\text{b}(\alpha_1-Y_1) \\ 
&-b_1\varepsilon_{\text{f}}\overline{X}_{1}^\text{b}\frac{\overline{C}_{3}}{\overline{C}_{3} + K_{31}} \\
&- \varepsilon_{\text{f}}\hat{q}^{\max}_{2}\frac{\overline{C}_{2}}{\overline{C}_{2} + K_2}\frac{\overline{C}_{3}}{\overline{C}_{3} + K_{32}}\overline{X}_{2}^\text{b}(\alpha_2-Y_2) \\&-b_2\varepsilon_{\text{f}}\overline{X}_{2}^\text{b}\frac{\overline{C}_{3}}{\overline{C}_{3} + K_{32}},
\end{align} 
\end{subequations}

\noindent where $p_i$ is the inactivation rate constant for species $i$, and $\alpha_i$ is the conversion factor for the reaction between species $i$ and substrate $C_3$.

\section{Governing equations for Case 3}\label{Eqs_Sec3}

\begin{subequations}
\begin{align}
\frac{\partial (\varepsilon_{\text{f}} \overline{X}_{1}^\text{sb})}{\partial t}
+ \nabla \mathbf{\cdot} (\overline{\mathbf{u}}_{\text{f}}\overline{X}_{1}^\text{sb})
- \nabla \cdot (\varepsilon_{\text{f}} D^*_{1} \nabla \overline{X}_{1}^\text{sb})
 &= \varepsilon_{\text{f}}Y_1\hat{q}^{\max}_{1}\frac{\overline{C}_{1}}{\overline{C}_{1} + K_1}\overline{X}_{1}^\text{sb}\\
 &+ \varepsilon_{\text{f}} k^{\text{det}} \tau^{\text{det}} \overline{X}_{1}^\text{b}
 - \varepsilon_{\text{f}} b_1 \overline{X}_{1}^\text{sb}, 
\end{align}
\begin{align}
\frac{\partial (\varepsilon_{\text{f}} \overline{X}_{1}^\text{b})}{\partial t}
 = \varepsilon_{\text{f}}Y_1\hat{q}^{\max}_{1}\frac{\overline{C}_{1}}{\overline{C}_{1} + K_1}\overline{X}_{1}^\text{b}
 - \varepsilon_{\text{f}} k^{\text{det}} \tau^{\text{det}} \overline{X}_{1}^\text{b}
 - \varepsilon_{\text{f}} b_1 \overline{X}_{1}^\text{b}, 
\end{align}
\begin{align}
\frac{\partial (\varepsilon_{\text{f}} \overline{C}_{1})}{\partial t}
+ \nabla \mathbf{\cdot} (\overline{\mathbf{u}}_{\text{f}}\overline{C}_{1})
- \nabla \cdot (\varepsilon_{\text{f}} D^*_{1} \nabla \overline{C}_{1})= - \varepsilon_{\text{f}}(\hat{q}^{\max}_{1}+b_1)\frac{\overline{C}_{1}}{\overline{C}_{1} + K_1}\left(\overline{X}_{1}^\text{sb}+\overline{X}_{1}^\text{b}\right).
\end{align} 
\end{subequations}

\section{Governing equations for counter-diffusional biofilms}
\label{sec:counter}
\begin{subequations}
\begin{align}
\frac{\partial (\varepsilon_{\text{f}} \overline{X}_{1}^\text{sb})}{\partial t}
+ \nabla \mathbf{\cdot}(\overline{\mathbf{u}}_{\text{f}}\overline{X}_{1}^{\text{sb}}) - \nabla \mathbf{\cdot} (\varepsilon_{\text{f}}D^*_{1} \nabla \overline{X}_{1}^{\text{sb}})
 &= \varepsilon_{\text{f}}Y_1\hat{q}^{\max}_{1}\frac{\overline{C}_{1}}{\overline{C}_{1} + K_1}\frac{\overline{C}_{3}}{\overline{C}_{3} + K_{3}}\overline{X}_{1}^\text{sb}\\
 &+\varepsilon_{\text{f}}k^{\text{det}}\tau^{\text{det}}\overline{X}_{1}^\text{b}
 -\varepsilon_{\text{f}}b_1\overline{X}_{1}^\text{sb},
\end{align}
\begin{align}
\frac{\partial (\varepsilon_{\text{f}} \overline{X}_{2}^\text{sb})}{\partial t}+ \nabla \mathbf{\cdot} (\overline{\mathbf{u}}_{\text{f}}\overline{X}_{2}^{\text{sb}}) - \nabla \mathbf{\cdot} (\varepsilon_{\text{f}}D^*_{2} \nabla \overline{X}_{2}^{\text{sb}})
 &= \varepsilon_{\text{f}}Y_2\hat{q}^{\max}_{2}\frac{\overline{C}_{2}}{\overline{C}_{2} + K_2}\frac{\overline{C}_{4}}{\overline{C}_{4} + K_{4}}\left( \frac{K_{32, \text{in}}}{\overline{C}_3 + K_{32, \text{in}}} \right)^{100}\overline{X}_{2}^\text{sb} \\
 &+\varepsilon_{\text{f}}k^{\text{det}}\tau^{\text{det}}\overline{X}_{2}^\text{b}
 -\varepsilon_{\text{f}}b_2\overline{X}_{2}^\text{sb},
\end{align}

\begin{align}
\frac{\partial (\varepsilon_{\text{f}} \overline{X}_{1}^\text{b})}{\partial t}
 = \varepsilon_{\text{f}}Y_1\hat{q}^{\max}_{1}\frac{\overline{C}_{1}}{\overline{C}_{1} + K_1}\frac{\overline{C}_{3}}{\overline{C}_{3} + K_{3}}\overline{X}_{1}^\text{b}
 -\varepsilon_{\text{f}}k^{\text{det}}\tau^{\text{det}}\overline{X}_{1}^\text{b}
 -\varepsilon_{\text{f}}b_1\overline{X}_{1}^\text{b},
\end{align}
\begin{align}
\frac{\partial (\varepsilon_{\text{f}} \overline{X}_{2}^\text{b})}{\partial t}
 = \varepsilon_{\text{f}}Y_2\hat{q}^{\max}_{2}\frac{\overline{C}_{2}}{\overline{C}_{2} + K_2}\frac{\overline{C}_{4}}{\overline{C}_{4} + K_{4}}\left( \frac{K_{32, \text{in}}}{\overline{C}_3 + K_{32, \text{in}}} \right)^{100}\overline{X}_{2}^\text{b}
 -\varepsilon_{\text{f}}k^{\text{det}}\tau^{\text{det}}\overline{X}_{2}^\text{b}
 -\varepsilon_{\text{f}}b_2\overline{X}_{2}^\text{b},
\end{align}

\begin{align}
\frac{\partial (\varepsilon_{\text{f}} \overline{C}_{1})}{\partial t}
+ \nabla \mathbf{\cdot} (\overline{\mathbf{u}}_{\text{f}}\overline{C}_{1})
- \nabla \cdot (\varepsilon_{\text{f}} D^*_{1} \nabla \overline{C}_{1})= - \varepsilon_{\text{f}}\hat{q}^{\max}_{1}\frac{\overline{C}_{1}}{\overline{C}_{1} + K_1}\frac{\overline{C}_{3}}{\overline{C}_{3} + K_{3}}(\overline{X}_{1}^\text{sb}+\overline{X}_{1}^\text{b}),
\end{align} 
\begin{align}
\frac{\partial (\varepsilon_{\text{f}} \overline{C}_{2})}{\partial t}
&+ \nabla \mathbf{\cdot} (\overline{\mathbf{u}}_{\text{f}}\overline{C}_2)
- \nabla \cdot (\varepsilon_{\text{f}} D^*_{2} \nabla \overline{C}_{2})\\
&= - \varepsilon_{\text{f}}\hat{q}^{\max}_{2}\frac{\overline{C}_{2}}{\overline{C}_{2} + K_2}\frac{\overline{C}_{4}}{\overline{C}_{4} + K_{4}}\left( \frac{K_{32, \text{in}}}{\overline{C}_3 + K_{32, \text{in}}} \right)^{100}
\left(\overline{X}_{2}^\text{sb}+\overline{X}_{2}^\text{b}\right)\\
&+\varepsilon_{\text{f}}\hat{q}^{\max}_{1}\frac{\overline{C}_{1}}{\overline{C}_{1} + K_1}\frac{\overline{C}_{3}}{\overline{C}_{3} + K_{3}}(\overline{X}_{1}^\text{sb}+\overline{X}_{1}^\text{b}),
\end{align} 
\begin{align}
\frac{\partial (\varepsilon_{\text{f}} \overline{C}_{3})}{\partial t}
+ \nabla \mathbf{\cdot} (\overline{\mathbf{u}}_{\text{f}}\overline{C}_{3})
- \nabla \cdot (\varepsilon_{\text{f}} D^*_{3} \nabla \overline{C}_{3})= - \varepsilon_{\text{f}}\hat{q}^{\max}_{1}\frac{\overline{C}_{1}}{\overline{C}_{1} + K_1}\frac{\overline{C}_{3}}{\overline{C}_{3} + K_{3}}(\overline{X}_{1}^\text{sb}+\overline{X}_{1}^\text{b}),
\end{align} 
\begin{align}
\frac{\partial (\varepsilon_{\text{f}} \overline{C}_{4})}{\partial t}
&+ \nabla \mathbf{\cdot} (\overline{\mathbf{u}}_{\text{f}}\overline{C}_4)
- \nabla \cdot (\varepsilon_{\text{f}} D^*_{4} \nabla \overline{C}_{4})\\
&= - \varepsilon_{\text{f}}\hat{q}^{\max}_{2}\frac{\overline{C}_{2}}{\overline{C}_{2} + K_2}\frac{\overline{C}_{4}}{\overline{C}_{4} + K_{4}}\left( \frac{K_{32, \text{in}}}{\overline{C}_3 + K_{32, \text{in}}} \right)^{100}
\left(\overline{X}_{2}^\text{sb}+\overline{X}_{2}^\text{b}\right),
\end{align} 
\end{subequations}

\bibliographystyle{elsarticle-num}


\begin{thebibliography}{10}
\expandafter\ifx\csname url\endcsname\relax
  \def\url#1{\texttt{#1}}\fi
\expandafter\ifx\csname urlprefix\endcsname\relax\def\urlprefix{URL }\fi
\expandafter\ifx\csname href\endcsname\relax
  \def\href#1#2{#2} \def\path#1{#1}\fi

\bibitem{Xian2022}
Y.~Xian, M.~Jin, H.~Zhan, X.~Liang, Permeable biofilms can support persistent hyporheic anoxic microzones, Geophysical Research Letters 49~(14) (2022) e2021GL096948.

\bibitem{Xian2019-lo}
Y.~Xian, M.~Jin, H.~Zhan, Y.~Liu, {Reactive transport of nutrients and bioclogging during dynamic disconnection process of stream and groundwater}, Water resources research 55~(5) (2019) 3882--3903.

\bibitem{Jasper2014-oo}
J.~T. Jasper, Z.~L. Jones, J.~O. Sharp, D.~L. Sedlak, {Nitrate removal in shallow, open-water treatment wetlands}, Environmental science \& technology 48~(19) (2014) 11512--11520.

\bibitem{Fang2017-yr}
H.~W. Fang, H.~J. Lai, W.~Cheng, L.~Huang, G.~J. He, {Modeling sediment transport with an integrated view of the biofilm effects: MODELING SEDIMENT TRANSPORT WITH {BIOFILM}}, Water Resources Research 53~(9) (2017) 7536--7557.

\bibitem{Feng2013}
L.~Feng, Z.~Wu, X.~Yu, Quorum sensing in water and wastewater treatment biofilms, Journal of environmental biology 34 (2013) 437--44.

\bibitem{Yuan2020}
K.~Yuan, S.~Li, F.~Zhong, Treatment of coking wastewater in biofilm-based bioaugmentation process: Biofilm formation and microbial community analysis, Journal of Hazardous Materials 400 (2020) 123117.

\bibitem{Lindholdt2015}
A.~Lindholdt, S.~Dam-Johansen, K.and~Olsen, D.~Yebra, S.~Kiil, Effects of biofouling development on drag forces of hull coatings for ocean-going ships: a review, Journal of Coatings Technology and Research 12 (2015) 415–444.

\bibitem{Ghoul2016}
M.~Ghoul, S.~Mitri, The ecology and evolution of microbial competition, Trends in Microbiology 24~(10) (2016) 833--845.

\bibitem{Baveye1998}
P.~Baveye, P.~Vandevivere, B.~L. Hoyle, P.~C. DeLeo, D.~S. de~Lozada, Environmental impact and mechanisms of the biological clogging of saturated soils and aquifer materials, Critical Reviews in Environmental Science and Technology 28~(2) (1998) 123--191.

\bibitem{Horn2003}
H.~Horn, H.~Reiff, E.~Morgenroth, Simulation of growth and detachment in biofilm systems under defined hydrodynamic conditions, Biotechnology and Bioengineering 81~(5) (2003) 607--617.

\bibitem{Kurz2022}
D.~L. Kurz, E.~Secchi, F.~J. Carrillo, I.~C. Bourg, R.~Stocker, J.~Jimenez-Martinez, Competition between growth and shear stress drives intermittency in preferential flow paths in porous medium biofilms, Proceedings of the National Academy of Sciences 119~(30) (2022) e2122202119.

\bibitem{Kurz2023}
D.~L. Kurz, E.~Secchi, R.~Stocker, J.~Jimenez-Martinez, Morphogenesis of biofilms in porous media and control on hydrodynamics, Environmental Science \& Technology 57~(14) (2023) 5666--5677.

\bibitem{Yan2017}
Z.~Yan, C.~Liu, Y.~Liu, V.~L. Bailey, Multiscale investigation on biofilm distribution and its impact on macroscopic biogeochemical reaction rates, Water Resources Research 53~(11) (2017) 8698--8714.

\bibitem{Deng2013}
W.~Deng, M.~B. Cardenas, M.~F. Kirk, S.~J. Altman, P.~C. Bennett, Effect of permeable biofilm on micro- and macro-scale flow and transport in bioclogged pores, Environ. Sci. Technol. 47 (2013) 11092--11098.

\bibitem{Vandevivere1992}
P.~Vandevivere, P.~Baveye, Saturated hydraulic conductivity reduction caused by aerobic bacteria in sand columns, Soil Science Society of America Journal 56~(1) (1992) 1--13.

\bibitem{Le-Bihan2000-ze}
Y.~Le~Bihan, P.~Lessard, {Monitoring biofilter clogging: biochemical characteristics of the biomass}, Water Research 34~(17) (2000) 4284--4294.

\bibitem{Jeong2017-br}
S.~Jeong, K.~Cho, D.~Jeong, S.~Lee, T.~Leiknes, S.~Vigneswaran, H.~Bae, {Effect of engineered environment on microbial community structure in biofilter and biofilm on reverse osmosis membrane}, Water Research 124 (2017) 227--237.

\bibitem{Mattei2017}
M.~R. Mattei, L.~Frunzo, B.~D'Acunto, Y.~P{\'e}chaud, F.~Pirozzi, G.~Esposito, Continuum and discrete approach in modeling biofilm development and structure: a review, Journal of Mathematical Biology 76 (2017) 945 -- 1003.

\bibitem{DAcunto2019}
B.~D'Acunto, L.~Frunzo, V.~Luongo, M.~R. Mattei, {Mathematical modeling of biofilms}, Vol.~5, IWA Publishing, 2006.

\bibitem{Eberl2001}
H.~J. Eberl, D.~F. Parker, M.~C. Van~Loosdrecht, A new deterministic spatio-temporal continuum model for biofilm development, Computational and Mathematical Methods in Medicine 3~(3) (2001) 161--175.

\bibitem{Picioreanu1998a}
C.~Picioreanu, van Loosdrecht~MC, H.~JJ, A new combined differential-discrete cellular automaton approach for biofilm modeling: application for growth in gel beads, Biotechnology and Bioengineering 57 (1998) 718--731.

\bibitem{Wanner1986}
O.~Wanner, W.~Gujer, A multispecies biofilm model, Biotechnology and Bioengineering 28~(3) (1986) 314--328.

\bibitem{Li2025}
X.~Li, X.~Yang, Pore-scale modeling of biofilm formation and biofilm-induced anomalous transport features in heterogenous porous media, Advances in Water Resources 195 (2025) 104877.

\bibitem{Benioug2017}
M.~Benioug, F.~Golfier, C.~Oltéan, M.~Buès, T.~Bahar, J.~Cuny, An immersed boundary-lattice boltzmann model for biofilm growth in porous media, Advances in Water Resources 107 (2017) 65--82.

\bibitem{Tang2013-dd}
Y.~Tang, A.~J. Valocchi, C.~J. Werth, H.~Liu, {An improved pore-scale biofilm model and comparison with a microfluidic flow cell experiment}, Water resources research 49~(12) (2013) 8370--8382.

\bibitem{Jung2021}
H.~Jung, C.~Meile, Pore-scale numerical investigation of evolving porosity and permeability driven by biofilm growth, Transport in Porous Media 139 (2021) 203--221.

\bibitem{Yousefzadeh2023-rg}
M.~Yousefzadeh, Y.~Yao, I.~Battiato, {A level-set immersed boundary method for reactive transport in complex topologies with moving interfaces}, Journal of computational physics 478~(111958) (2023) 111958.

\bibitem{Li2010-ro}
X.~Li, H.~Huang, P.~Meakin, {A three-dimensional level set simulation of coupled reactive transport and precipitation/dissolution}, International journal of heat and mass transfer 53~(13) (2010) 2908--2923.

\bibitem{Li2008-bs}
X.~Li, H.~Huang, P.~Meakin, {Level set simulation of coupled advection-diffusion and pore structure evolution due to mineral precipitation in porous media}, Water resources research 44~(12) (Dec. 2008).

\bibitem{Alpkvista2007}
E.~Alpkvista, I.~Klapper, A multidimensional multispecies continuum model for heterogeneous biofilm development, Bulletin of Mathematical Biology 69~(2) (2007) 765--789.

\bibitem{Picioreanu1998b}
C.~Picioreanu, van Loosdrecht~MC, H.~JJ, Mathematical modeling of biofilm structure with a hybrid differential- discrete cellular automaton approach, Biotechnology and Bioengineering 58 (1998) 101--116.

\bibitem{Tang2013}
Y.~Tang, A.~J. Valocchi, An improved cellular automaton method to model multispecies biofilms, Water Research 47~(15) (2013) 5729--5742.

\bibitem{Rahman2015}
K.~A. Rahman, R.~Sudarsan, H.~J. Eberl, A mixed-culture biofilm model with cross-diffusion, Bulletin of Mathematical Biology 77~(11) (2015) 2086--2124.

\bibitem{Wood2002}
B.~D. Wood, M.~Quintard, S.~Whitaker, Calculation of effective diffusivities for biofilms and tissues, Biotechnology and Bioengineering 77~(5) (2002) 495--516.

\bibitem{Golfier2009}
F.~Golfier, B.~D. Wood, L.~Orgogozo, M.~Quintard, M.~Buès, Biofilms in porous media: Development of macroscopic transport equations via volume averaging with closure for local mass equilibrium conditions, Advances in Water Resources 32~(3) (2009) 463--485.

\bibitem{Klapper2002}
I.~Klapper, J.~Dockery, Finger formation in biofilm layers, SIAM Journal on Applied Mathematics 62~(3) (2002) 853--869.

\bibitem{Cockx2024}
B.~J. Cockx, T.~Foster, R.~J. Clegg, K.~Alden, S.~Arya, D.~J. Stekel, B.~F. Smets, J.-U. Kreft, Is it selfish to be filamentous in biofilms? individual-based modeling links microbial growth strategies with morphology using the new and modular idynomics 2.0, PLoS Computational Biology 20~(2) (2024) e1011303.

\bibitem{Pizarro2001}
G.~Pizarro, D.~Griffeath, D.~Noguera, Quantitative cellular automaton model for biofilms, Journal of Environmental Engineering-asce - J ENVIRON ENG-ASCE 127 (2001) 782 -- 789.

\bibitem{Noguera2004}
D.~Noguera, C.~Picioreanu, Results from the multi-species benchmark problem 3 (bm3) using two-dimensional models, Water Science and Technology 49~(11-12) (2004) 169--176.

\bibitem{Kreft2001}
J.-U. Kreft, C.~Picioreanu, J.~W.~T. Wimpenny, M.~C.~M. van Loosdrecht, Individual-based modelling of biofilms, Microbiology 147~(11) (2001) 2897--2912.

\bibitem{Noguera1999}
D.~R. Noguera, G.~Pizarfo, D.~A. Stahl, B.~E. Rittmann, Simulation of multispecies biofilm development in three dimensions, Water Science and Technology 39~(7) (1999) 123--130, microbial ecology of biofilms.

\bibitem{Knutson2005}
C.~E. Knutson, C.~J. Werth, A.~J. Valocchi, Pore-scale simulation of biomass growth along the transverse mixing zone of a model two-dimensional porous medium, Water Resources Research 41~(7) (2005).

\bibitem{Laspidou2010}
C.~Laspidou, A.~Kungolos, P.~Samaras, Cellular-automata and individual-based approaches for the modeling of biofilm structures: Pros and cons, Desalination 250~(1) (2010) 390--394.

\bibitem{Thullner2008}
M.~Thullner, P.~Baveye, Computational pore network modeling of the influence of biofilm permeability on bioclogging in porous media, Biotechnology and Bioengineering 99~(6) (2008) 1337--1351.

\bibitem{Stoodley1994}
P.~Stoodley, D.~DeBeer, Z.~Lewandowski, Liquid flow in biofilm systems, Applied and environmental microbiology 60~(8) (1994) 2711--2716.

\bibitem{Dreszer2013}
C.~Dreszer, J.~Vrouwenvelder, A.~Paulitsch-Fuchs, A.~Zwijnenburg, J.~Kruithof, H.-C. Flemming, Hydraulic resistance of biofilms, Journal of Membrane Science 429 (2013) 436--447.

\bibitem{Karimifard2021}
S.~Karimifard, X.~Li, C.~Elowsky, Y.~Li, Modeling the impact of evolving biofilms on flow in porous media inside a microfluidic channel, Water Research 188 (2021) 116536.

\bibitem{Golfier2015}
F.~Golfier, D.~Lasseux, M.~Quintard, Investigation of the effective permeability of vuggy or fractured porous media from a darcy-brinkman approach, Computational Geosciences 19 (2015) 63--78.

\bibitem{Soulaine2018}
C.~Soulaine, S.~Roman, A.~Kovscek, H.~A. Tchelepi, Pore-scale modelling of multiphase reactive flow: application to mineral dissolution with production of $\text{CO}_{2}$, Journal of Fluid Mechanics 855 (2018) 616–645.

\bibitem{Guo2018}
B.~Guo, L.~Ma, H.~A. Tchelepi, Image-based micro-continuum model for gas flow in organic-rich shale rock, Advances in Water Resources 122 (2018) 70--84.

\bibitem{Carrillo2022}
F.~J. Carrillo, C.~Soulaine, I.~C. Bourg, The impact of sub-resolution porosity on numerical simulations of multiphase flow, Advances in Water Resources 161 (2022) 104094.

\bibitem{Soulaine2016}
C.~Soulaine, H.~A. Tchelepi, Micro-continuum approach for pore-scale simulation of subsurface processes, Transport in Porous Media 113 (2016) 431–456.

\bibitem{Guibert2025}
R.~Guibert, P.~Horgue, G.~Debenest, Micro-continuum modeling of biofilm growth coupled with hydrodynamics in openfoam, SoftwareX 29 (2025) 102011.

\bibitem{Kapellos2007-qs}
G.~E. Kapellos, T.~S. Alexiou, S.~Pavlou, A.~C. Payatakes, {Hierarchical simulation of biofilm growth dynamics in porous media}, Environmental Science and Technology, American Science Press, Houston, USA 1 (2007) 491--502.

\bibitem{Vafai1981}
K.~Vafai, C.~Tien, Boundary and inertia effects on flow and heat transfer in porous media, International Journal of Heat and Mass Transfer 24~(2) (1981) 195--203.

\bibitem{Hsu1990}
C.~Hsu, P.~Cheng, Thermal dispersion in a porous medium, International Journal of Heat and Mass Transfer 33~(8) (1990) 1587--1597.

\bibitem{Bousquet2002}
P.~Bousquet-Melou, B.~Goyeau, M.~Quintard, F.~Fichot, D.~Gobin, Average momentum equation for interdendritic flow in a solidifying columnar mushy zone, International Journal of Heat and Mass Transfer 45~(17) (2002) 3651--3665.

\bibitem{Goyeau2003}
B.~Goyeau, D.~Lhuillier, D.~Gobin, M.~Velarde, Momentum transport at a fluid–porous interface, International Journal of Heat and Mass Transfer 46~(21) (2003) 4071--4081.

\bibitem{Lewandowski1999}
Z.~Lewandowski, D.~Webb, M.~Hamilton, G.~Harkin, Quantifying biofilm structure, Water Science and Technology 39~(7) (1999) 71--76.

\bibitem{Zhang1994}
T.~C. Zhang, P.~L. Bishop, Density, porosity, and pore structure of biofilms, Water Research 28~(11) (1994) 2267--2277.

\bibitem{Zang2025}
G.~Zang, L.~Huang, S.~Wang, T.~Lu, Y.~Gong, L.~Chen, Study on the effects of non-uniformity of microbial growth on permeability changes in porous media, Advances in Water Resources 195 (2025) 104876.

\bibitem{bear2013dynamics}
J.~Bear, Dynamics of fluids in porous media, Courier Corporation, 2013.

\bibitem{Zhu1995}
W.~Zhu, C.~David, T.-f. Wong, Network modeling of permeability evolution during cementation and hot isostatic pressing, Journal of Geophysical Research: Solid Earth 100~(B8) (1995) 15451--15464.

\bibitem{Hommel2018}
J.~Hommel, E.~Coltman, H.~Class, Porosity–permeability relations for evolving pore space: A review with a focus on (bio-)geochemically altered porous media, Transport in Porous Media 124 (2018) 589–629.

\bibitem{kozeny1927}
J.~Kozeny, Ueber kapillare leitung des wassers im boden, Sitzungsberichte der Akademie der Wissenschaften in Wien 136 (1927) 271.

\bibitem{carman1937}
P.~C. Carman, Fluid flow through granular beds, Trans. Inst. Chem. Eng. London 15 (1937) 150--156.

\bibitem{Minto2019}
J.~M. Minto, R.~J. Lunn, G.~El~Mountassir, Development of a reactive transport model for field-scale simulation of microbially induced carbonate precipitation, Water Resources Research 55~(8) (2019) 7229--7245.

\bibitem{ISSA1986}
R.~Issa, Solution of the implicitly discretised fluid flow equations by operator-splitting, Journal of Computational Physics 62~(1) (1986) 40--65.

\bibitem{Lardon2011}
L.~A. Lardon, B.~V. Merkey, S.~Martins, A.~D\"{o}tsch, C.~Picioreanu, J.-U. Kreft, B.~F. Smets, idynomics: next-generation individual-based modelling of biofilms, Environmental Microbiology 13~(9) (2011) 2416--2434.

\bibitem{Mehrani2021}
M.-J. Mehrani, X.~Lu, P.~Kowal, D.~Sobotka, J.~M\c{a}kinia, Incorporation of the complete ammonia oxidation (comammox) process for modeling nitrification in suspended growth wastewater treatment systems, Journal of Environmental Management 297 (2021) 113223.

\bibitem{Wang2022}
Z.~Wang, P.~Perez-Calleja, R.~Nerenberg, P.~J. Novak, S.~Ishii, Unraveling encapsulated growth of nitrosomonas europaea in alginate: An experimental and modeling study, Water Research 208 (2022) 117857.

\bibitem{Wang2020}
Z.~Wang, V.~E. Pane, C.~S. Criddle, Metabolic model of nitrite reduction to nitrous oxide coupled to alternating consumption and storage of glycogen and polyhydroxyalkanoate, Bioresource Technology Reports 9 (2020) 100370.

\bibitem{Guanju2023}
G.~Wei, J.~Yang, Impacts of hydrodynamic conditions and microscale surface roughness on the critical shear stress to develop and thickness of early‐stage pseudomonas putida biofilms, Biotechnology and Bioengineering 120 (04 2023).

\bibitem{Choi2003}
Y.~Choi, E.~Morgenroth, Monitoring biofilm detachment under dynamic changes in shear stress using laser-based particle size analysis and mass fractionation, Water Science and Technology 47~(5) (2003) 69--76.

\bibitem{Stoodley2001}
P.~Stoodley, S.~Wilson, L.~Hall-Stoodley, J.~D. Boyle, H.~M. Lappin-Scott, J.~W. Costerton, Growth and detachment of cell clusters from mature mixed-species biofilms, Applied and Environmental Microbiology 67~(12) (2001) 5608--5613.

\bibitem{Sudarsan2005}
R.~Sudarsan, K.~Milferstedt, E.~Morgenroth, H.~Eberl, Quantification of detachment forces on rigid biofilm colonies in a roto-torque reactor using computational fluid dynamics tools, Water Science and Technology 52~(7) (2005) 149--154.

\bibitem{Nerenberg2016-qx}
R.~Nerenberg, {The membrane-biofilm reactor (MBfR) as a counter-diffusional biofilm process}, Current opinion in biotechnology 38 (2016) 131--136.

\bibitem{Martin2015-vn}
K.~J. Martin, C.~Picioreanu, R.~Nerenberg, {Assessing microbial competition in a hydrogen-based membrane biofilm reactor (MBfR) using multidimensional modeling: Modeling Competition in a Hydrogen-Based, Membrane Biofilm Reactor}, Biotechnology and bioengineering 112~(9) (2015) 1843--1853.

\bibitem{Shukla2026-is}
R.~Shukla, Z.~Wang, M.~K. Jeon, T.~Yan, {Iron as an electron reservoir in two-zone biofilters for robust nitrogen removal}, ACS ES\&T Engineering 6~(3) (2026) 1160--1171.

\bibitem{Shanahan2004}
J.~Shanahan, M.~J. Semmens, Multipopulation model of membrane-aerated biofilms, Environmental science \& technology 38 (2004) 3176--83.

\end{thebibliography}

\end{document}